\documentclass[aps,prl,reprint,superscriptaddress,nofootinbib,floatfix]{revtex4-2}
\usepackage{graphicx}
\usepackage{dcolumn}
\usepackage{bm}

\usepackage{multirow}

\usepackage{amssymb}
\usepackage{amsmath}
\usepackage{graphicx}
\usepackage{xcolor}
\usepackage{braket}
\usepackage{CJK}
\usepackage{indentfirst}
\usepackage{amsmath}
\usepackage{cases}
\usepackage[T1]{fontenc}

\def\Eq#1{Eq.~(\ref{#1})}
\def\Fig#1{Fig.~\ref{#1}}

\def\Inc#1{\left[#1\right]}
\newcommand{\ra}{\rightarrow}

\renewcommand{\[}{\left[}

\newcommand{\beq} {\begin{equation}}
\newcommand{\eeq} {\end{equation}}
\newcommand{\bea} {\begin{eqnarray}}
\newcommand{\eea} {\end{eqnarray}}

\definecolor{UkiyoRed}{RGB}{223,126,102}
\definecolor{UkiyoGreen}{RGB}{148,181,148}
\definecolor{UkiyoYellow}{RGB}{237,199,117}
\definecolor{YinshuaiBlue}{RGB}{130,143,199}
\definecolor{YinshuaiPurple}{RGB}{134,128,207}
\definecolor{YinshuaiOrange}{RGB}{255,203,148}
\usepackage{hyperref}
\hypersetup{colorlinks=true, citecolor=blue, urlcolor=blue, linkcolor=blue}

\begin{document}
\title{High-temperature charge-$4e$ superconductivity in SU(4) interacting fermions}

\author{Shao-Hang Shi}
\affiliation{Beijing National Laboratory for Condensed Matter Physics \& Institute of Physics, Chinese Academy of Sciences, Beijing 100190, China}
\affiliation{University of Chinese Academy of Sciences, Beijing 100049, China}
\author{Zhengzhi Wu}
\affiliation{Rudolf Peierls Centre for Theoretical Physics, Parks Road, Oxford, OX1 3PU, UK}
\author{Jiangping Hu}\email{jphu@iphy.ac.cn}
\affiliation{Beijing National Laboratory for Condensed Matter Physics \& Institute of Physics, Chinese Academy of Sciences, Beijing 100190, China}
\affiliation{University of Chinese Academy of Sciences, Beijing 100049, China}
	\affiliation{New Cornerstone Science Laboratory,  Institute of Physics, Chinese Academy of Sciences, Beijing 100190, China}
\author{Zi-Xiang Li}
\email{zixiangli@iphy.ac.cn}
\affiliation{Beijing National Laboratory for Condensed Matter Physics \& Institute of Physics, Chinese Academy of Sciences, Beijing 100190, China}
\affiliation{University of Chinese Academy of Sciences, Beijing 100049, China}

\date{\today}
\maketitle

\textbf{The condensation of electron quartets, known as charge-$4e$ superconductivity (SC), represents a novel quantum state of matter beyond the standard paradigm of Cooper pairing. However, concrete microscopic models realizing this phase in two dimensions remain a central challenge. Here,  we introduce a non-engineered and sign-problem-free model, unambiguously demonstrating the emergence of a robust and high-temperature charge-$4e$ SC phase by unbiased quantum Monte-Carlo. At zero temperature, the phase diagram reveals that charge-$4e$ SC is the primary ground state in the strong-coupling regime. At finite temperature in the absence of charge-$2e$ SC, we identify charge-$4e$ SC through a Berezinskii-Kosterlitz-Thouless transition, marked by a universal jump in the superfluid stiffness consistent with a condensate of charge $4e$. Remarkably, the transition temperature $T_c$ increases nearly linearly with interaction strength,  providing a robust mechanism for high-$T_c$ quartet SC. Furthermore, spectral analysis reveals a prominent pseudogap above $T_c$ arising from strong phase fluctuations.  Our results establish a canonical and numerically exact model system for charge-$4e$ SC, offering crucial guidance for its realization in experimental platforms such as moir\'e materials and ultracold atomic systems.}

\section{Introduction}
 Unraveling unconventional superconductivity (SC) driven by novel mechanisms beyond the BCS paradigm is a central topic of enduring interest in modern condensed matter physics. Charge-$4e$ SC, characterized by the condensation of electron quartets rather than traditional Cooper pairs, is a novel quantum phase fundamentally distinct from the conventional superconducting state~\cite{Wu2005PRLQuartet,Kivelson1990PRB,Berg2009NP,Jiang2017PRBQMCCharge4e,Jiang2024SB,Yang2023PRBCharge4e,Cui2023PRL,Gao2025arXivcharge4e,Shi2026arXivcharge4e}. The quest to observe charge-$4e$ SC has attracted tremendous experimental interest for several decades; however, despite some preliminary signatures~\cite{Wang2024PRX,Kim2022NC,Cohen2018PNAS}, unambiguous experimental evidence remains scarce. Theoretically, the charge-$4e$ state is conventionally proposed to emerge as a vestigial order after primary orders—such as pair-density-wave~\cite{Zhou2022NCCharge4eKagome,Wu2024npjQM} or nematic superconductivity~\cite{Jian2021PRLNematicCharge4e,Fu2021PRLNematicCharge4e}—are destroyed by fluctuations~\cite{Berg2009njp,Yang2023NC,Fernandes2023PRB,Yao2020PRLHolsteinHubbard,Fernandes2024PRBNematiccharge4e,Varma2023PRBCharge4e,Vishwanath2022PRBCharge4e,Yao2025arXivcharge4e,Zhang2025PRB,Zhang2025arXiv}. It has also been predicted to occur in one-dimensional interacting systems~\cite{Titus2024PRB} and, more recently, in moir\'e materials~\cite{Wu2024PRLWTe2,Song2025arXiv}. Nevertheless, microscopic models that explicitly realize charge-$4e$ SC in dimensions higher than one are rare~\cite{Wang2022PRBSYK,Chirolli2024PRR,Gao2026arXivcharge4e1}. In particular, an unambiguous demonstration of charge-4e SC in a microscopic electronic model in two or more dimensions using unbiased methods is still lacking.

\begin{figure}[t]
\includegraphics[width= 0.95\linewidth]{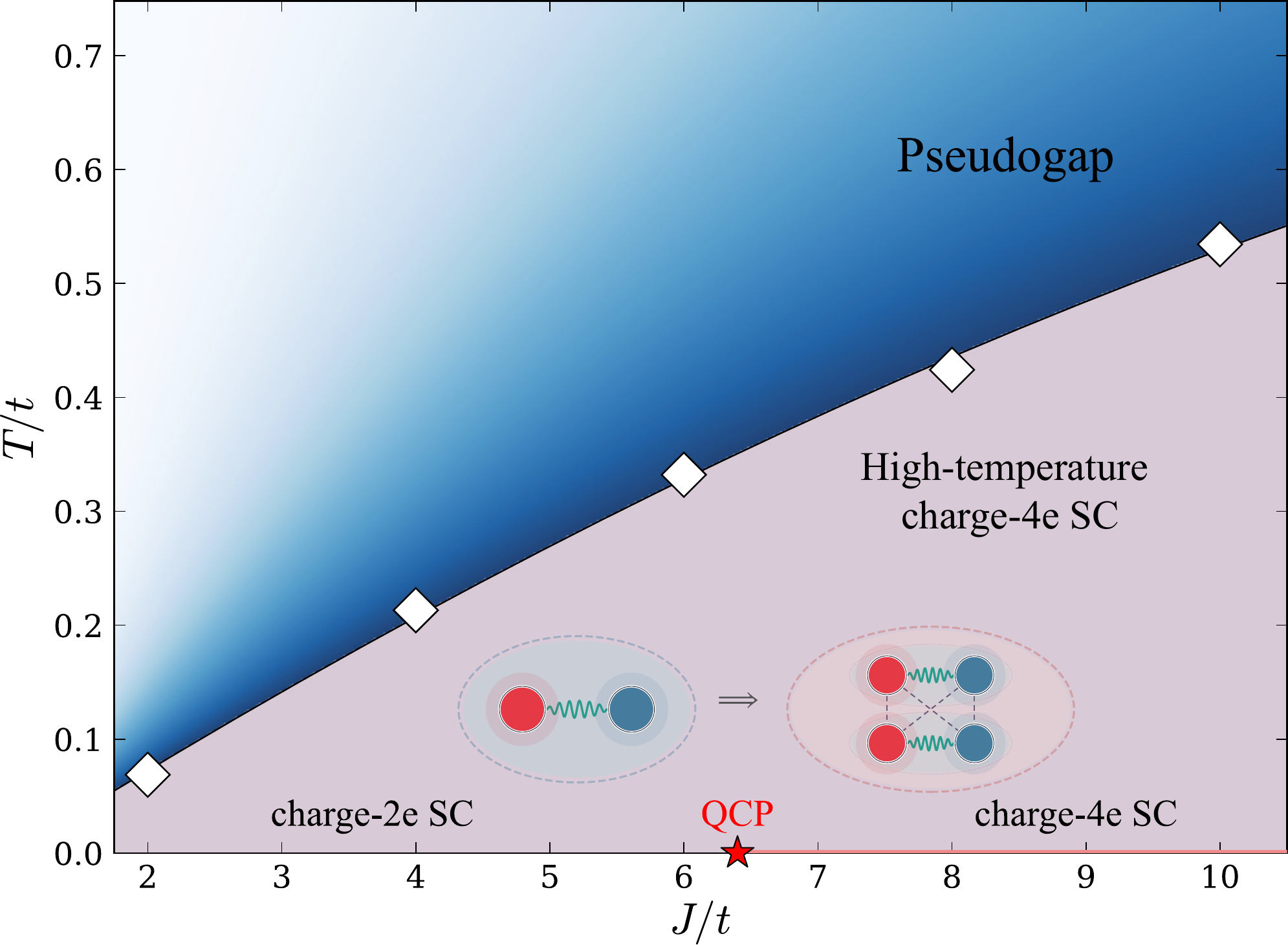}~~	
\caption{{\bf Sketch of the phase diagram.} At zero temperature, charge-$2e$ superconductivity (SC) dominates in the weak-coupling regime, while charge-$4e$ SC emerges as the primary order at strong coupling. At finite temperature, the charge-$2e$ pairing remains strictly short-ranged. In contrast, the transition temperature $T_c$ of the charge-$4e$ SC exhibits a nearly linear increase with interaction strength in the strong-coupling regime. Above $T_c$, strong superconducting fluctuations give rise to a prominent pseudogap region. }
\label{Fig1}
\end{figure}

In this Letter, we bridge this critical gap by demonstrating the emergence of charge-$4e$ SC in an SU(4) interacting electronic model using sign-problem-free quantum Monte Carlo (QMC) simulations~\cite{li2019review,AssaadReview}. Specifically, we systematically investigate a doped SU(4) fermionic model with Su-Schrieffer-Heeger (SSH)-type interactions, whose half-filled ground state hosts quantum spin liquid (QSL) and various valence-bond-solid (VBS) phases. Our main results are summarized as follows: 
\textbf{(1)} Unbiased QMC calculations reveal that robust SC emerges upon doping the QSL and VBS ground states. In the weak-coupling regime, the ground state is an on-site charge-$2e$ SC, whereas in the strong-coupling regime, it transitions into a primary charge-$4e$ SC. 
\textbf{(2)} At any finite temperature, the charge-$2e$ SC order vanishes, while the charge-$4e$ SC persists over a wide temperature range, characterized by a finite superfluid stiffness. The universal jump in the superfluid stiffness to $8T_c/\pi$ provides compelling evidence for a charge-$4e$ Berezinskii-Kosterlitz-Thouless (BKT) transition. Notably, the transition temperature $T_c$ increases nearly linearly with the interaction strength, yielding a high-$T_c$ charge-$4e$ SC at strong coupling. 
\textbf{(3)} At temperatures above $T_c$, strong superconducting fluctuations give rise to distinct pseudogap features. 
Collectively, our results provide a robust theoretical platform for investigating charge-$4e$ SC and offer crucial guidance for its experimental realization.

\section{Results}
\subsection{Model and Method}
We consider a microscopic model of SU($N$) fermions on a square lattice with SSH-type interactions~\cite{Li2017NC,Cai2021PRLSSH,Li2024PRLSU4}, described by the Hamiltonian:
\begin{equation}
\label{E1}
\hat{H} = -t\sum_{\langle ij \rangle,\alpha} (\hat{c}^{\dagger}_{i\alpha}\hat{c}_{j\alpha} + \mathrm{H.c.}) - \frac{J}{2N}\sum_{\langle ij \rangle} \left(\sum_\alpha \hat{c}^{\dagger}_{i\alpha}\hat{c}_{j\alpha} + \mathrm{H.c.}\right)^{2},
\end{equation}
where $\langle ij \rangle$ denotes nearest-neighbor (NN) bonds, and $c^{\dagger}_{i\alpha}$ creates a fermion at site $i$ with flavor index $\alpha=1,\dots,N$. Hereafter, we focus on the $N=4$ case and set the hopping amplitude $t=1$ as the unit of energy. The parameter $J$ dictates the strength of the SSH interaction, which effectively arises from electron-phonon coupling (EPC) in the antiadiabatic (fast-phonon) limit~\cite{Fradkin1983PRB,Cai2021PRLSSH}. At half-filling, the ground-state phase diagram of this model was established in our previous work~\cite{Li2024PRLSU4}. Depending on the interaction strength, two distinct VBS orders emerge: a staggered VBS in the weakly interacting regime ($J < 1.5$) and a columnar VBS in the strongly interacting regime ($J > 5.0$). In the intermediate regime, competition between these two orders stabilizes a gapped QSL. Remarkably, for even $N$, the Hamiltonian in Eq.~(\ref{E1}) is free from the notorious sign problem at generic fillings in determinant quantum Monte Carlo (DQMC) simulations~\cite{li2019review,Wu2005PRB,Li2016PRL,Troyer2005sign}. It thus provides a highly promising platform to investigate the unconventional superconductivity that may emerge upon doping the QSL or VBS phases.

In this work, we employ unbiased, large-scale projector DQMC to study the ground-state properties of the model~\cite{AssaadReview,Sorella1989EPL}, alongside finite-temperature DQMC to investigate its thermal behavior~\cite{AssaadReview,BSS}, with details regarding the QMC simulations provided in the Supplementary Materials (SM). To characterize the SC orders, we compute the equal-time structure factor: 
\begin{equation}
S(\mathbf{q},L) = \frac{1}{L^{4}}\sum_{i,j}e^{i\mathbf{q}\cdot(\mathbf{R}_{i}-\mathbf{R}_{j})}\langle \hat{O}^{\dagger}_{i} \hat{O}_{j}\rangle,
\label{E2}
\end{equation} 
where $L$ is the linear system size and the ordering momentum is $\mathbf{q}=(0,0)$. The dominant channel for the charge-$2e$ SC is on-site $s$-wave pairing, defined by $\hat{O}_{i} = \hat{c}_{i\alpha}\hat{c}_{i\beta}$; without loss of generality, we take $\hat{O}_{i} = \hat{c}_{i1}\hat{c}_{i2}$. For the charge-$4e$ SC, the local electron quartet operator is defined as $\hat{O}_{i} = \hat{c}_{i1}\hat{c}_{i2}\hat{c}_{i3}\hat{c}_{i4}$.  To unambiguously identify true long-range order, we evaluate the correlation ratio:
\begin{equation}
R(L) = 1 - \frac{1}{4}\sum_{\delta\mathbf{q}_{\mathrm{min}}}\frac{S(\mathbf{Q}+\delta\mathbf{q}_{\mathrm{min}},L) }{S(\mathbf{Q},L)} ,
\end{equation}
where $\mathbf{Q}$ maximizes the structure factor and $\delta\mathbf{q}_{\mathrm{min}} = \{ (\pm\frac{2\pi}{L}, 0), (0, \pm\frac{2\pi}{L}) \}$ is the minimal lattice momentum step. In a long-range ordered phase, $R(L)$ increases with $L$, whereas it decreases with $L$ if the order is short-range. Consequently, $R(L)$ provides a powerful metric for determining the onset of long-range order and precisely locating phase transition points.

The phase stiffness of Cooper pairs against fluctuations is a fundamental property of superconductivity, directly governing the Meissner effect and zero resistivity. To characterize this macroscopic phase coherence, we compute the superfluid stiffness $\rho_s$~\cite{Zhang1993PRB}:
\bea
\rho_s=\frac{1}{4}\Inc{-K_x-\Lambda_{xx}(q_x=0,q_y\ra 0,\omega_n=0)},
\label{E4}
\eea
where $K_x$ denotes the diamagnetic current response along the $x$ direction, with  details included in the SM. $\Lambda_{xx}(\mathbf{q},\omega_n)$ is the paramagnetic current correlator defined as: $
\Lambda_{xx}(\mathbf{q}, i\omega_n) =  \int_{0}^{\beta} d\tau \, e^{i\omega_n\tau} 
 \langle \hat{J_x}(\mathbf{q}, \tau) \hat{J_x}(-\mathbf{q}, 0) \rangle
$, where $\hat{J_x}(\mathbf{q}, \tau)$ represents the current operator in the $\hat{x}$-direction with momentum $\mathbf{q}$ and imaginary time $\tau$.

\begin{figure*}[t]
\includegraphics[width= 1.0\linewidth]{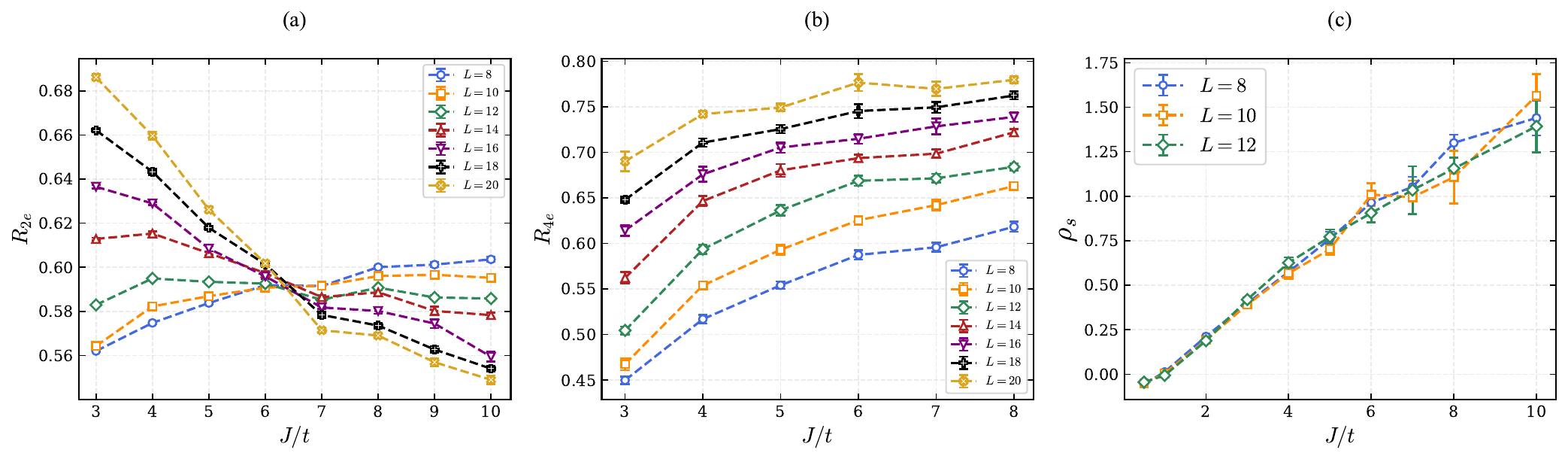}~~	
\caption{ \textbf{Evidence for robust primary charge-$4e$ superconductivity in the strongly interacting ground state.} The doping level is fixed at $\delta = 0.15$. 
\textbf{(a)} Correlation-length ratio $R_{\mathrm{2e}}(L)$ for charge-$2e$ on-site $s$-wave superconductivity as a function of interaction strength $J$, shown for various linear system sizes $L$. The crossing point at $J_c\approx 6.3$ marks a quantum phase transition separating the charge-$2e$ long-range and short-range ordered phases. 
\textbf{(b)} Charge-$4e$ correlation-length ratio $R_{\mathrm{4e}}(L)$ versus $L$ for different interaction strengths. The monotonic increase of $R_{\mathrm{4e}}(L)$ with system size across the entire interaction regime provides unambiguous evidence for long-range order. 
\textbf{(c)} Superfluid stiffness $\rho_s$ as a function of $J$. The finite value of $\rho_s$ confirms macroscopic phase coherence, while its monotonic increase indicates that the superconducting state becomes more robust in the strong-coupling regime. }
\label{Fig2}
\end{figure*}

\subsection{Primary charge-$4e$ SC at strong interaction}

 To investigate the emergence of SC upon doping the VBS and QSL phases, we fix the doping level at $\delta=0.15$ and tune the interaction strength $J$. We begin by examining the ground-state charge-$2e$ SC. In \Fig{Fig2}(a), we present the correlation-length ratio $R_{\rm 2e}(L)$ for the on-site charge-$2e$ pairing. In the weak- and intermediate-coupling regimes, $R_{\rm 2e}(L)$ increases with system size $L$, demonstrating the presence of long-range charge-$2e$ superconducting order. However, as $J$ increases further, $R_{\rm 2e}(L)$ drops rapidly for sufficiently large $L$, yielding a well-defined size-independent crossing point at $J_c \approx 6.3$. In the strong-coupling regime ($J > J_c$), $R_{\rm 2e}(L)$ decreases with $L$, indicating that the charge-$2e$ SC order becomes strictly short-ranged.

In contrast, for the charge-$4e$ SC, the correlation-length ratio $R_{\rm 4e}(L)$, as shown in \Fig{Fig2}(b), monotonically increases with both $J$ and $L$ across the entire interaction range considered. Consequently, in the strong-coupling regime ($J \geq J_c$), the charge-$4e$ order is long-ranged while the charge-$2e$ SC is strictly short-ranged, establishing the charge-$4e$ SC as the primary order. Conversely, in the weak-coupling regimes, charge-$2e$ pairing remains dominant due to conventional Cooper instabilities at the Fermi surface. Thus, as the interaction strength increases, the system undergoes a distinct quantum phase transition from a charge-$2e$ on-site $s$-wave SC to a primary charge-$4e$ SC. The finite-size scaling of the structure factors for both orders further substantiates this conclusion, as detailed in the SM.



To further verify the superconducting nature of the ground state, we evaluate the superfluid stiffness $\rho_s$, a key quantity characterizing macroscopic phase coherence. As presented in \Fig{Fig2}(c), $\rho_s$ is intrinsically finite and exhibits negligible finite-size effects, directly confirming a true superconducting ground state. Furthermore, the superfluid stiffness increases monotonically with the interaction strength $J$, indicating that superconductivity remains highly robust even in the strongly interacting regime. Overall, these findings demonstrate that robust SC emerges upon doping the VBS or QSL phases, directly corroborating the long-standing paradigm of superconductivity arising from a doped resonating valence bond (RVB) state~\cite{Anderson1987Science}. Most crucially, by combining the scaling behaviors of $R_{\rm 2e}(L)$ and $R_{\rm 4e}(L)$, our data unambiguously establish the emergence of a primary charge-$4e$ SC upon doping the SU(4) fermions with SSH interaction in strongly coupling regime.  

\begin{figure*}[htbp]
    \centering   
    \includegraphics[width=\textwidth]{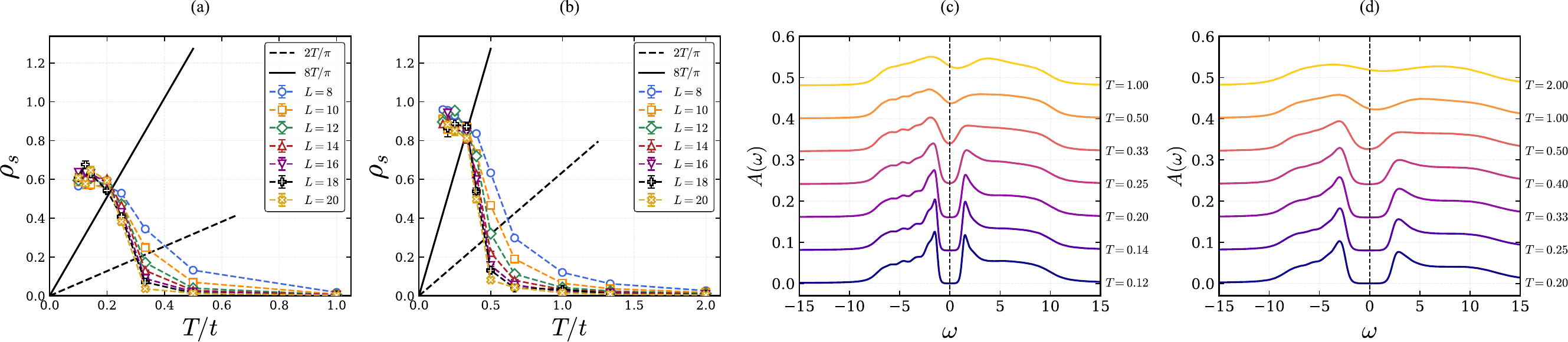}
    \caption{
\textbf{Finite-temperature signatures of charge-$4e$ superconductivity and pseudogap formation.} The doping level is fixed at $\delta = 0.15$.
\textbf{(a)}, \textbf{(b)} Superfluid stiffness $\rho_s$ as a function of temperature $T$ for interaction strengths $J=4.0$ (a) and $J=6.0$ (b). The abrupt drop to zero signifies a BKT transition. The solid line in each panel represents the universal relation $\rho_s(T_c^-) = (8/\pi)T_c$, whereas the dash line denotes the relation $\rho_s(T) = (2/\pi)T$.  The intersection of solid line with the $\rho_s(T)$ curve indicates $T_c$, which aligns with the abrupt drop in $\rho_s$. This agreement confirms the charge-$4e$ nature of the superconducting state. From these data, we extract $T_c = 0.23$ for $J=4.0$, $T_c = 0.32$ for $J=6.0$.
\textbf{(c)}, \textbf{(d)}: Single-particle spectral function $A(\omega)$ for $J=4.0$ (c) and $J=6.0$ (d). At low temperatures ($T < T_c$), a hard superconducting gap is visible. Crucially, a distinct double-peak structure, indicating a pseudogap, persists in a wide temperature regime above the superconducting transition ($T_c < T < T^*$). This demonstrates the survival of preformed electron quartets without long-range phase coherence, a hallmark of a pseudogap originating from strong phase fluctuations. 
    }
    \label{Fig3}
\end{figure*}

\subsection{ High-temperature charge-$4e$ superconductivity}

At finite temperatures, the charge-$2e$ SC order parameter spontaneously breaks the continuous non-Abelian SU(4) flavor symmetry. According to the Mermin-Wagner theorem, any long-range or quasi-long-range order associated with this broken SU(4) symmetry must be strictly destroyed by thermal fluctuations at any non-zero temperature. On the other hand, the charge-$4e$ SC state is formed by an SU(4) singlet of electron quartets. Because it only breaks the Abelian charge U(1) symmetry, quasi-long-range order of the charge-$4e$ SC can persist at finite temperatures, driven by a BKT mechanism. In this BKT phase, algebraic correlations survive as long as vortex-antivortex pairs remain bound. As the temperature increases up to the BKT transition temperature, thermal fluctuations cause these vortices to unbind and proliferate, ultimately destroying macroscopic phase coherence and the charge-$4e$ quasi-long-range order. 

For a two-dimensional system at non-zero temperatures, a finite superfluid stiffness provides direct evidence for the existence of superconducting quasi-long-range order. We present the calculated superfluid stiffness, defined in \Eq{E4}, in \Fig{Fig3}. The results explicitly demonstrate that a finite superfluid stiffness persists over a broad temperature range, confirming the survival of macroscopic phase coherence. Because charge-$2e$ SC is strictly forbidden at finite temperatures in this SU(4)-symmetric system, the emergent superconducting phase observed in this thermal regime should therefore originate from charge-$4e$ SC. A prominent characteristic of charge-$4e$ SC is that its fundamental vortex excitations carry a fractional magnetic flux of $h/4e$, exactly one quarter of the elementary flux quantum~\cite{Berg2009NP}. Because the BKT transition depends on the square of the condensate charge, this fractional flux yields a direct, observable consequence: the universal jump in the superfluid stiffness at $T_c$ becomes $\rho_s(T_c^-) = \frac{8}{\pi} T_c$, in stark contrast to the $\frac{2}{\pi} T_c$ jump characteristic of conventional charge-$2e$ pairing~\cite{BKT1972,KT1973,KT1974,Lee1985PRB}. The numerical results presented in \Fig{Fig3} explicitly demonstrate that the jump in the superfluid stiffness aligns precisely with the $\frac{8}{\pi} T_c$ relation, rather than $\frac{2}{\pi} T_c$. These findings provide unambiguous evidence for the emergence of charge-$4e$ SC in the doped SU(4) fermionic model with SSH interactions at finite temperature.

By identifying the temperature at which the superfluid stiffness undergoes this universal jump, we extract the BKT transition temperature $T_c$ of the charge-$4e$ SC state. Remarkably, $T_c$ increases nearly linearly with the interaction strength $J$, reaching an exceptionally high value of $0.5t$ at $J=10.0$ within the parameter regime considered. The dependence of $T_c$ on the interaction strength $J$ is summarized in the finite-temperature phase diagram in \Fig{Fig1}. This continued enhancement of $T_c$ in the strongly interacting regime stems from the specific nature of the SSH-type interaction, which effectively arises from SSH EPC in the anti-adiabatic limit. For standard on-site interactions or Holstein EPC, the effective mass of the electron pairs increases exponentially at strong coupling; this heavily suppresses macroscopic phase coherence and leads to a sharp reduction in $T_c$. In contrast, as discussed in recent works on SU(2) systems~\cite{Cai2025PRB,Zhang2023PRXBipolaron}, because SSH-type phonons couple directly to electron hopping, they generate effective pair-hopping terms that actively enhance phase coherence in the strong coupling regime. In our SU(4) model, an analogous mechanism governs the strong-interaction limit, sustaining robust phase coherence and ultimately giving rise to high-$T_c$ charge-$4e$ SC.

We evaluate the spectral functions $A(\omega)$ at various temperatures for different interaction strengths $J$ using stochastic analytical continuation~\cite{sandvik1998SAC} of the imaginary-time Green's function from our QMC simulations. As the temperature increases, the spectral function at the Fermi energy, $A(\omega=0)$, rises sharply at $T_c$, an effect we attribute to vortex proliferation, consistent with recent ARPES observations in cuprate superconductors~\cite{ZXShen2022Nature,He2021PRXSCfluctuation}. In the intermediate and strong interaction regimes, a double-peak feature, manifesting a spectral gap at finite temperature, persists over a wide temperature range above $T_c$. The particle-hole symmetric nature of this feature indicates that the pseudogap arises from phase fluctuations~\cite{Kivelson1995Nature}. Consequently, our numerical results reveal that a charge-$4e$ superconducting phase with strong fluctuation exists in doped SU(4) systems with SSH interactions, particularly in the strong coupling regime.

\section{Concluding remarks}
 In this Letter, we present the first unbiased numerical demonstration of a primary charge-$4e$ superconducting ground state in a 2D microscopic model. Our large-scale QMC simulations convincingly demonstrate that stable SC emerges upon doping the QSL and VBS insulating phases of an SU(4) fermionic model with SSH interactions. As the interaction strength increases, the system undergoes a quantum phase transition from an on-site charge-$2e$ SC to a primary charge-$4e$ SC. Remarkably, the $T_c$ continues to rise even deep within the strong-coupling regime. Our unbiased results thus establish a compelling theoretical platform for exploring primary and high-$T_c$ charge-$4e$ SC.

Our unbiased numerical results provide a promising route toward the experimental realization of charge-$4e$ SC. In particular, because the $T_c$ is significantly enhanced in the strong-coupling regime, high-$T_c$ charge-$4e$ superconductivity could potentially be realized in SU(4) fermionic systems. Although exact SU(4) symmetry is often broken in realistic solid-state materials, SU(4)-symmetric interacting fermion systems have been successfully engineered in cold-atom optical lattices~\cite{Yoshiro2010PRLSUN,Sengstock2012NPSUN,Fallani2014NPSUN,Yoshiro2012NPSU6,Wu2018PRLSUN,Wu2024npjQMReview,Matsuda2014Review}. Furthermore, 2D moir\'e materials offer an effective SU(4)-symmetric platform by combining multiple degrees of freedom, such as spin and valley~\cite{Xu2018PRL,Zhang2021PRLSU4,Classen2023PRB}. In effective models describing twisted bilayer graphene, the dominant EPC is bond-type, representing a generalized SSH interaction. The nearly flat bands in these moir\'e systems naturally facilitate both the strong-coupling and antiadiabatic limits, making them ideal candidates to host this exotic phase. Finally, while generic SU(4) symmetry-breaking terms in quantum materials may allow charge-$2e$ SC to emerge at finite temperatures,  we expect the primary charge-$4e$ SC to remain robust over a broad temperature window, provided the symmetry breaking is not dominant. We leave a systematic study of this interplay to future work. Consequently, our findings offer an important starting point for the experimental search for charge-$4e$ SC in realistic systems.

\subsection{Acknowledgments} 

We sincerely thank Hong Yao, Zheng-Yu Weng, Hui Yang, Zhi-Qiang Gao, Yan-Qi Wang and Zhou-Quan Wan for helpful discussions. This work is supported by the National Natural Science Foundation of China under Grant Nos. 12347107 and 12474146, Beijing Natural Science Foundation under Grant No. JR25007. J.H. and Z.-X.L. are supported by the New Cornerstone Investigator Program. Z.W. acknowledges the support from the EPSRC under grant EP/X030881/1.

{\em Note added}: Upon completing this work, we became aware of an interesting related work by Z.-Q. Wan, H. Jiang, X. Zou, S. Zhang and S.-K. Jian, which appeared on arXiv around the same time as our work and reported the primary charge-$4e$ superconductivity in the SU(4) attractive Hubbard model~\cite{Jiancharge4e}. The model they studied and primary focus in their work differ from ours.

\bibliography{ref}

\begin{thebibliography}{68}%
\makeatletter
\providecommand \@ifxundefined [1]{%
 \@ifx{#1\undefined}
}%
\providecommand \@ifnum [1]{%
 \ifnum #1\expandafter \@firstoftwo
 \else \expandafter \@secondoftwo
 \fi
}%
\providecommand \@ifx [1]{%
 \ifx #1\expandafter \@firstoftwo
 \else \expandafter \@secondoftwo
 \fi
}%
\providecommand \natexlab [1]{#1}%
\providecommand \enquote  [1]{``#1''}%
\providecommand \bibnamefont  [1]{#1}%
\providecommand \bibfnamefont [1]{#1}%
\providecommand \citenamefont [1]{#1}%
\providecommand \href@noop [0]{\@secondoftwo}%
\providecommand \href [0]{\begingroup \@sanitize@url \@href}%
\providecommand \@href[1]{\@@startlink{#1}\@@href}%
\providecommand \@@href[1]{\endgroup#1\@@endlink}%
\providecommand \@sanitize@url [0]{\catcode `\\12\catcode `\$12\catcode `\&12\catcode `\#12\catcode `\^12\catcode `\_12\catcode `\%12\relax}%
\providecommand \@@startlink[1]{}%
\providecommand \@@endlink[0]{}%
\providecommand \url  [0]{\begingroup\@sanitize@url \@url }%
\providecommand \@url [1]{\endgroup\@href {#1}{\urlprefix }}%
\providecommand \urlprefix  [0]{URL }%
\providecommand \Eprint [0]{\href }%
\providecommand \doibase [0]{https://doi.org/}%
\providecommand \selectlanguage [0]{\@gobble}%
\providecommand \bibinfo  [0]{\@secondoftwo}%
\providecommand \bibfield  [0]{\@secondoftwo}%
\providecommand \translation [1]{[#1]}%
\providecommand \BibitemOpen [0]{}%
\providecommand \bibitemStop [0]{}%
\providecommand \bibitemNoStop [0]{.\EOS\space}%
\providecommand \EOS [0]{\spacefactor3000\relax}%
\providecommand \BibitemShut  [1]{\csname bibitem#1\endcsname}%
\let\auto@bib@innerbib\@empty
\bibitem [{\citenamefont {Wu}(2005)}]{Wu2005PRLQuartet}%
  \BibitemOpen
  \bibfield  {author} {\bibinfo {author} {\bibfnamefont {C.}~\bibnamefont {Wu}},\ }\bibfield  {title} {\bibinfo {title} {Competing orders in one-dimensional spin-$3/2$ fermionic systems},\ }\href {https://doi.org/10.1103/PhysRevLett.95.266404} {\bibfield  {journal} {\bibinfo  {journal} {Phys. Rev. Lett.}\ }\textbf {\bibinfo {volume} {95}},\ \bibinfo {pages} {266404} (\bibinfo {year} {2005})}\BibitemShut {NoStop}%
\bibitem [{\citenamefont {Kivelson}\ \emph {et~al.}(1990)\citenamefont {Kivelson}, \citenamefont {Emery},\ and\ \citenamefont {Lin}}]{Kivelson1990PRB}%
  \BibitemOpen
  \bibfield  {author} {\bibinfo {author} {\bibfnamefont {S.~A.}\ \bibnamefont {Kivelson}}, \bibinfo {author} {\bibfnamefont {V.~J.}\ \bibnamefont {Emery}},\ and\ \bibinfo {author} {\bibfnamefont {H.~Q.}\ \bibnamefont {Lin}},\ }\bibfield  {title} {\bibinfo {title} {Doped antiferromagnets in the weak-hopping limit},\ }\href {https://doi.org/10.1103/PhysRevB.42.6523} {\bibfield  {journal} {\bibinfo  {journal} {Phys. Rev. B}\ }\textbf {\bibinfo {volume} {42}},\ \bibinfo {pages} {6523} (\bibinfo {year} {1990})}\BibitemShut {NoStop}%
\bibitem [{\citenamefont {Berg}\ \emph {et~al.}(2009{\natexlab{a}})\citenamefont {Berg}, \citenamefont {Fradkin},\ and\ \citenamefont {Kivelson}}]{Berg2009NP}%
  \BibitemOpen
  \bibfield  {author} {\bibinfo {author} {\bibfnamefont {E.}~\bibnamefont {Berg}}, \bibinfo {author} {\bibfnamefont {E.}~\bibnamefont {Fradkin}},\ and\ \bibinfo {author} {\bibfnamefont {S.~A.}\ \bibnamefont {Kivelson}},\ }\bibfield  {title} {\bibinfo {title} {Charge-4e superconductivity from pair-density-wave order in certain high-temperature superconductors},\ }\href {https://doi.org/10.1038/nphys1389} {\bibfield  {journal} {\bibinfo  {journal} {Nature Physics}\ }\textbf {\bibinfo {volume} {5}},\ \bibinfo {pages} {830} (\bibinfo {year} {2009}{\natexlab{a}})}\BibitemShut {NoStop}%
\bibitem [{\citenamefont {Jiang}\ \emph {et~al.}(2017)\citenamefont {Jiang}, \citenamefont {Li}, \citenamefont {Kivelson},\ and\ \citenamefont {Yao}}]{Jiang2017PRBQMCCharge4e}%
  \BibitemOpen
  \bibfield  {author} {\bibinfo {author} {\bibfnamefont {Y.-F.}\ \bibnamefont {Jiang}}, \bibinfo {author} {\bibfnamefont {Z.-X.}\ \bibnamefont {Li}}, \bibinfo {author} {\bibfnamefont {S.~A.}\ \bibnamefont {Kivelson}},\ and\ \bibinfo {author} {\bibfnamefont {H.}~\bibnamefont {Yao}},\ }\bibfield  {title} {\bibinfo {title} {Charge-$4e$ superconductors: A majorana quantum monte carlo study},\ }\href {https://doi.org/10.1103/PhysRevB.95.241103} {\bibfield  {journal} {\bibinfo  {journal} {Phys. Rev. B}\ }\textbf {\bibinfo {volume} {95}},\ \bibinfo {pages} {241103} (\bibinfo {year} {2017})}\BibitemShut {NoStop}%
\bibitem [{\citenamefont {Li}\ \emph {et~al.}(2024)\citenamefont {Li}, \citenamefont {Jiang},\ and\ \citenamefont {Hu}}]{Jiang2024SB}%
  \BibitemOpen
  \bibfield  {author} {\bibinfo {author} {\bibfnamefont {P.}~\bibnamefont {Li}}, \bibinfo {author} {\bibfnamefont {K.}~\bibnamefont {Jiang}},\ and\ \bibinfo {author} {\bibfnamefont {J.}~\bibnamefont {Hu}},\ }\bibfield  {title} {\bibinfo {title} {Charge 4e superconductor: A wavefunction approach},\ }\href {https://doi.org/https://doi.org/10.1016/j.scib.2024.06.002} {\bibfield  {journal} {\bibinfo  {journal} {Science Bulletin}\ }\textbf {\bibinfo {volume} {69}},\ \bibinfo {pages} {2328} (\bibinfo {year} {2024})}\BibitemShut {NoStop}%
\bibitem [{\citenamefont {Qin}\ \emph {et~al.}(2023)\citenamefont {Qin}, \citenamefont {Dong}, \citenamefont {Sheng}, \citenamefont {Huang},\ and\ \citenamefont {Yang}}]{Yang2023PRBCharge4e}%
  \BibitemOpen
  \bibfield  {author} {\bibinfo {author} {\bibfnamefont {Q.}~\bibnamefont {Qin}}, \bibinfo {author} {\bibfnamefont {J.-J.}\ \bibnamefont {Dong}}, \bibinfo {author} {\bibfnamefont {Y.}~\bibnamefont {Sheng}}, \bibinfo {author} {\bibfnamefont {D.}~\bibnamefont {Huang}},\ and\ \bibinfo {author} {\bibfnamefont {Y.-f.}\ \bibnamefont {Yang}},\ }\bibfield  {title} {\bibinfo {title} {Superconducting fluctuations and charge-4$e$ plaquette state at strong coupling},\ }\href {https://doi.org/10.1103/PhysRevB.108.054506} {\bibfield  {journal} {\bibinfo  {journal} {Phys. Rev. B}\ }\textbf {\bibinfo {volume} {108}},\ \bibinfo {pages} {054506} (\bibinfo {year} {2023})}\BibitemShut {NoStop}%
\bibitem [{\citenamefont {Liu}\ \emph {et~al.}(2023{\natexlab{a}})\citenamefont {Liu}, \citenamefont {Wang},\ and\ \citenamefont {Cui}}]{Cui2023PRL}%
  \BibitemOpen
  \bibfield  {author} {\bibinfo {author} {\bibfnamefont {R.}~\bibnamefont {Liu}}, \bibinfo {author} {\bibfnamefont {W.}~\bibnamefont {Wang}},\ and\ \bibinfo {author} {\bibfnamefont {X.}~\bibnamefont {Cui}},\ }\bibfield  {title} {\bibinfo {title} {Quartet superfluid in two-dimensional mass-imbalanced fermi mixtures},\ }\href {https://doi.org/10.1103/PhysRevLett.131.193401} {\bibfield  {journal} {\bibinfo  {journal} {Phys. Rev. Lett.}\ }\textbf {\bibinfo {volume} {131}},\ \bibinfo {pages} {193401} (\bibinfo {year} {2023}{\natexlab{a}})}\BibitemShut {NoStop}%
\bibitem [{\citenamefont {Gao}\ \emph {et~al.}(2026{\natexlab{a}})\citenamefont {Gao}, \citenamefont {Wang}, \citenamefont {Yang},\ and\ \citenamefont {Wu}}]{Gao2025arXivcharge4e}%
  \BibitemOpen
  \bibfield  {author} {\bibinfo {author} {\bibfnamefont {Z.-Q.}\ \bibnamefont {Gao}}, \bibinfo {author} {\bibfnamefont {Y.-Q.}\ \bibnamefont {Wang}}, \bibinfo {author} {\bibfnamefont {H.}~\bibnamefont {Yang}},\ and\ \bibinfo {author} {\bibfnamefont {C.}~\bibnamefont {Wu}},\ }\href {https://arxiv.org/abs/2512.21325} {\bibinfo {title} {Topological charge-2ne superconductors}} (\bibinfo {year} {2026}{\natexlab{a}}),\ \Eprint {https://arxiv.org/abs/2512.21325} {arXiv:2512.21325 [cond-mat.str-el]} \BibitemShut {NoStop}%
\bibitem [{\citenamefont {Shi}\ \emph {et~al.}(2026)\citenamefont {Shi}, \citenamefont {Han}, \citenamefont {Raghu},\ and\ \citenamefont {Vishwanath}}]{Shi2026arXivcharge4e}%
  \BibitemOpen
  \bibfield  {author} {\bibinfo {author} {\bibfnamefont {Z.~D.}\ \bibnamefont {Shi}}, \bibinfo {author} {\bibfnamefont {Z.}~\bibnamefont {Han}}, \bibinfo {author} {\bibfnamefont {S.}~\bibnamefont {Raghu}},\ and\ \bibinfo {author} {\bibfnamefont {A.}~\bibnamefont {Vishwanath}},\ }\href {https://arxiv.org/abs/2602.06963} {\bibinfo {title} {Charge-$4e$ superconductor with parafermionic vortices: A path to universal topological quantum computation}} (\bibinfo {year} {2026}),\ \Eprint {https://arxiv.org/abs/2602.06963} {arXiv:2602.06963 [cond-mat.str-el]} \BibitemShut {NoStop}%
\bibitem [{\citenamefont {Ge}\ \emph {et~al.}(2024)\citenamefont {Ge}, \citenamefont {Wang}, \citenamefont {Xing}, \citenamefont {Yin}, \citenamefont {Wang}, \citenamefont {Shen}, \citenamefont {Lei}, \citenamefont {Wang},\ and\ \citenamefont {Wang}}]{Wang2024PRX}%
  \BibitemOpen
  \bibfield  {author} {\bibinfo {author} {\bibfnamefont {J.}~\bibnamefont {Ge}}, \bibinfo {author} {\bibfnamefont {P.}~\bibnamefont {Wang}}, \bibinfo {author} {\bibfnamefont {Y.}~\bibnamefont {Xing}}, \bibinfo {author} {\bibfnamefont {Q.}~\bibnamefont {Yin}}, \bibinfo {author} {\bibfnamefont {A.}~\bibnamefont {Wang}}, \bibinfo {author} {\bibfnamefont {J.}~\bibnamefont {Shen}}, \bibinfo {author} {\bibfnamefont {H.}~\bibnamefont {Lei}}, \bibinfo {author} {\bibfnamefont {Z.}~\bibnamefont {Wang}},\ and\ \bibinfo {author} {\bibfnamefont {J.}~\bibnamefont {Wang}},\ }\bibfield  {title} {\bibinfo {title} {Charge-$4e$ and charge-$6e$ flux quantization and higher charge superconductivity in kagome superconductor ring devices},\ }\href {https://doi.org/10.1103/PhysRevX.14.021025} {\bibfield  {journal} {\bibinfo  {journal} {Phys. Rev. X}\ }\textbf {\bibinfo {volume} {14}},\ \bibinfo {pages} {021025} (\bibinfo {year} {2024})}\BibitemShut {NoStop}%
\bibitem [{\citenamefont {Huang}\ \emph {et~al.}(2022)\citenamefont {Huang}, \citenamefont {Ronen}, \citenamefont {M{\'e}lin}, \citenamefont {Feinberg}, \citenamefont {Watanabe}, \citenamefont {Taniguchi},\ and\ \citenamefont {Kim}}]{Kim2022NC}%
  \BibitemOpen
  \bibfield  {author} {\bibinfo {author} {\bibfnamefont {K.-F.}\ \bibnamefont {Huang}}, \bibinfo {author} {\bibfnamefont {Y.}~\bibnamefont {Ronen}}, \bibinfo {author} {\bibfnamefont {R.}~\bibnamefont {M{\'e}lin}}, \bibinfo {author} {\bibfnamefont {D.}~\bibnamefont {Feinberg}}, \bibinfo {author} {\bibfnamefont {K.}~\bibnamefont {Watanabe}}, \bibinfo {author} {\bibfnamefont {T.}~\bibnamefont {Taniguchi}},\ and\ \bibinfo {author} {\bibfnamefont {P.}~\bibnamefont {Kim}},\ }\bibfield  {title} {\bibinfo {title} {Evidence for 4e charge of cooper quartets in a biased multi-terminal graphene-based josephson junction},\ }\href {https://doi.org/10.1038/s41467-022-30732-7} {\bibfield  {journal} {\bibinfo  {journal} {Nature Communications}\ }\textbf {\bibinfo {volume} {13}},\ \bibinfo {pages} {3032} (\bibinfo {year} {2022})}\BibitemShut {NoStop}%
\bibitem [{\citenamefont {Cohen}\ \emph {et~al.}(2018)\citenamefont {Cohen}, \citenamefont {Ronen}, \citenamefont {Kang}, \citenamefont {Heiblum}, \citenamefont {Feinberg}, \citenamefont {M{\'e}lin},\ and\ \citenamefont {Shtrikman}}]{Cohen2018PNAS}%
  \BibitemOpen
  \bibfield  {author} {\bibinfo {author} {\bibfnamefont {Y.}~\bibnamefont {Cohen}}, \bibinfo {author} {\bibfnamefont {Y.}~\bibnamefont {Ronen}}, \bibinfo {author} {\bibfnamefont {J.-H.}\ \bibnamefont {Kang}}, \bibinfo {author} {\bibfnamefont {M.}~\bibnamefont {Heiblum}}, \bibinfo {author} {\bibfnamefont {D.}~\bibnamefont {Feinberg}}, \bibinfo {author} {\bibfnamefont {R.}~\bibnamefont {M{\'e}lin}},\ and\ \bibinfo {author} {\bibfnamefont {H.}~\bibnamefont {Shtrikman}},\ }\bibfield  {title} {\bibinfo {title} {Nonlocal supercurrent of quartets in a three-terminal josephson junction},\ }\href {https://doi.org/10.1073/pnas.1800044115} {\bibfield  {journal} {\bibinfo  {journal} {Proceedings of the National Academy of Sciences}\ }\textbf {\bibinfo {volume} {115}},\ \bibinfo {pages} {6991} (\bibinfo {year} {2018})}\BibitemShut {NoStop}%
\bibitem [{\citenamefont {Zhou}\ and\ \citenamefont {Wang}(2022)}]{Zhou2022NCCharge4eKagome}%
  \BibitemOpen
  \bibfield  {author} {\bibinfo {author} {\bibfnamefont {S.}~\bibnamefont {Zhou}}\ and\ \bibinfo {author} {\bibfnamefont {Z.}~\bibnamefont {Wang}},\ }\bibfield  {title} {\bibinfo {title} {Chern fermi pocket, topological pair density wave, and charge-4e and charge-6e superconductivity in kagom{\'e} superconductors},\ }\href {https://doi.org/10.1038/s41467-022-34832-2} {\bibfield  {journal} {\bibinfo  {journal} {Nature Communications}\ }\textbf {\bibinfo {volume} {13}},\ \bibinfo {pages} {7288} (\bibinfo {year} {2022})}\BibitemShut {NoStop}%
\bibitem [{\citenamefont {Wu}\ and\ \citenamefont {Wang}(2024)}]{Wu2024npjQM}%
  \BibitemOpen
  \bibfield  {author} {\bibinfo {author} {\bibfnamefont {Y.-M.}\ \bibnamefont {Wu}}\ and\ \bibinfo {author} {\bibfnamefont {Y.}~\bibnamefont {Wang}},\ }\bibfield  {title} {\bibinfo {title} {d-wave charge-4e superconductivity from fluctuating pair density waves},\ }\href {https://doi.org/10.1038/s41535-024-00674-y} {\bibfield  {journal} {\bibinfo  {journal} {npj Quantum Materials}\ }\textbf {\bibinfo {volume} {9}},\ \bibinfo {pages} {66} (\bibinfo {year} {2024})}\BibitemShut {NoStop}%
\bibitem [{\citenamefont {Jian}\ \emph {et~al.}(2021)\citenamefont {Jian}, \citenamefont {Huang},\ and\ \citenamefont {Yao}}]{Jian2021PRLNematicCharge4e}%
  \BibitemOpen
  \bibfield  {author} {\bibinfo {author} {\bibfnamefont {S.-K.}\ \bibnamefont {Jian}}, \bibinfo {author} {\bibfnamefont {Y.}~\bibnamefont {Huang}},\ and\ \bibinfo {author} {\bibfnamefont {H.}~\bibnamefont {Yao}},\ }\bibfield  {title} {\bibinfo {title} {Charge-$4e$ superconductivity from nematic superconductors in two and three dimensions},\ }\href {https://doi.org/10.1103/PhysRevLett.127.227001} {\bibfield  {journal} {\bibinfo  {journal} {Phys. Rev. Lett.}\ }\textbf {\bibinfo {volume} {127}},\ \bibinfo {pages} {227001} (\bibinfo {year} {2021})}\BibitemShut {NoStop}%
\bibitem [{\citenamefont {Fernandes}\ and\ \citenamefont {Fu}(2021)}]{Fu2021PRLNematicCharge4e}%
  \BibitemOpen
  \bibfield  {author} {\bibinfo {author} {\bibfnamefont {R.~M.}\ \bibnamefont {Fernandes}}\ and\ \bibinfo {author} {\bibfnamefont {L.}~\bibnamefont {Fu}},\ }\bibfield  {title} {\bibinfo {title} {Charge-$4e$ superconductivity from multicomponent nematic pairing: Application to twisted bilayer graphene},\ }\href {https://doi.org/10.1103/PhysRevLett.127.047001} {\bibfield  {journal} {\bibinfo  {journal} {Phys. Rev. Lett.}\ }\textbf {\bibinfo {volume} {127}},\ \bibinfo {pages} {047001} (\bibinfo {year} {2021})}\BibitemShut {NoStop}%
\bibitem [{\citenamefont {Berg}\ \emph {et~al.}(2009{\natexlab{b}})\citenamefont {Berg}, \citenamefont {Fradkin}, \citenamefont {Kivelson},\ and\ \citenamefont {Tranquada}}]{Berg2009njp}%
  \BibitemOpen
  \bibfield  {author} {\bibinfo {author} {\bibfnamefont {E.}~\bibnamefont {Berg}}, \bibinfo {author} {\bibfnamefont {E.}~\bibnamefont {Fradkin}}, \bibinfo {author} {\bibfnamefont {S.~A.}\ \bibnamefont {Kivelson}},\ and\ \bibinfo {author} {\bibfnamefont {J.~M.}\ \bibnamefont {Tranquada}},\ }\bibfield  {title} {\bibinfo {title} {Striped superconductors: how spin, charge and superconducting orders intertwine in the cuprates},\ }\href {https://doi.org/10.1088/1367-2630/11/11/115004} {\bibfield  {journal} {\bibinfo  {journal} {New Journal of Physics}\ }\textbf {\bibinfo {volume} {11}},\ \bibinfo {pages} {115004} (\bibinfo {year} {2009}{\natexlab{b}})}\BibitemShut {NoStop}%
\bibitem [{\citenamefont {Liu}\ \emph {et~al.}(2023{\natexlab{b}})\citenamefont {Liu}, \citenamefont {Zhou}, \citenamefont {Wu},\ and\ \citenamefont {Yang}}]{Yang2023NC}%
  \BibitemOpen
  \bibfield  {author} {\bibinfo {author} {\bibfnamefont {Y.-B.}\ \bibnamefont {Liu}}, \bibinfo {author} {\bibfnamefont {J.}~\bibnamefont {Zhou}}, \bibinfo {author} {\bibfnamefont {C.}~\bibnamefont {Wu}},\ and\ \bibinfo {author} {\bibfnamefont {F.}~\bibnamefont {Yang}},\ }\bibfield  {title} {\bibinfo {title} {Charge-4e superconductivity and chiral metal in 45{\textdegree}-twisted bilayer cuprates and related bilayers},\ }\href {https://doi.org/10.1038/s41467-023-43782-2} {\bibfield  {journal} {\bibinfo  {journal} {Nature Communications}\ }\textbf {\bibinfo {volume} {14}},\ \bibinfo {pages} {7926} (\bibinfo {year} {2023}{\natexlab{b}})}\BibitemShut {NoStop}%
\bibitem [{\citenamefont {Hecker}\ \emph {et~al.}(2023)\citenamefont {Hecker}, \citenamefont {Willa}, \citenamefont {Schmalian},\ and\ \citenamefont {Fernandes}}]{Fernandes2023PRB}%
  \BibitemOpen
  \bibfield  {author} {\bibinfo {author} {\bibfnamefont {M.}~\bibnamefont {Hecker}}, \bibinfo {author} {\bibfnamefont {R.}~\bibnamefont {Willa}}, \bibinfo {author} {\bibfnamefont {J.}~\bibnamefont {Schmalian}},\ and\ \bibinfo {author} {\bibfnamefont {R.~M.}\ \bibnamefont {Fernandes}},\ }\bibfield  {title} {\bibinfo {title} {Cascade of vestigial orders in two-component superconductors: Nematic, ferromagnetic, $s$-wave charge-$4e$, and $d$-wave charge-$4e$ states},\ }\href {https://doi.org/10.1103/PhysRevB.107.224503} {\bibfield  {journal} {\bibinfo  {journal} {Phys. Rev. B}\ }\textbf {\bibinfo {volume} {107}},\ \bibinfo {pages} {224503} (\bibinfo {year} {2023})}\BibitemShut {NoStop}%
\bibitem [{\citenamefont {Han}\ \emph {et~al.}(2020)\citenamefont {Han}, \citenamefont {Kivelson},\ and\ \citenamefont {Yao}}]{Yao2020PRLHolsteinHubbard}%
  \BibitemOpen
  \bibfield  {author} {\bibinfo {author} {\bibfnamefont {Z.}~\bibnamefont {Han}}, \bibinfo {author} {\bibfnamefont {S.~A.}\ \bibnamefont {Kivelson}},\ and\ \bibinfo {author} {\bibfnamefont {H.}~\bibnamefont {Yao}},\ }\bibfield  {title} {\bibinfo {title} {Strong coupling limit of the {Holstein-Hubbard} model},\ }\href {https://doi.org/10.1103/PhysRevLett.125.167001} {\bibfield  {journal} {\bibinfo  {journal} {Phys. Rev. Lett.}\ }\textbf {\bibinfo {volume} {125}},\ \bibinfo {pages} {167001} (\bibinfo {year} {2020})}\BibitemShut {NoStop}%
\bibitem [{\citenamefont {Hecker}\ and\ \citenamefont {Fernandes}(2024)}]{Fernandes2024PRBNematiccharge4e}%
  \BibitemOpen
  \bibfield  {author} {\bibinfo {author} {\bibfnamefont {M.}~\bibnamefont {Hecker}}\ and\ \bibinfo {author} {\bibfnamefont {R.~M.}\ \bibnamefont {Fernandes}},\ }\bibfield  {title} {\bibinfo {title} {Local condensation of charge-$4e$ superconductivity at a nematic domain wall},\ }\href {https://doi.org/10.1103/PhysRevB.109.134514} {\bibfield  {journal} {\bibinfo  {journal} {Phys. Rev. B}\ }\textbf {\bibinfo {volume} {109}},\ \bibinfo {pages} {134514} (\bibinfo {year} {2024})}\BibitemShut {NoStop}%
\bibitem [{\citenamefont {Varma}\ and\ \citenamefont {Wang}(2023)}]{Varma2023PRBCharge4e}%
  \BibitemOpen
  \bibfield  {author} {\bibinfo {author} {\bibfnamefont {C.~M.}\ \bibnamefont {Varma}}\ and\ \bibinfo {author} {\bibfnamefont {Z.}~\bibnamefont {Wang}},\ }\bibfield  {title} {\bibinfo {title} {Extended superconducting fluctuation region and $6e$ and $4e$ flux quantization in a kagome compound with a normal state of $3q$ order},\ }\href {https://doi.org/10.1103/PhysRevB.108.214516} {\bibfield  {journal} {\bibinfo  {journal} {Phys. Rev. B}\ }\textbf {\bibinfo {volume} {108}},\ \bibinfo {pages} {214516} (\bibinfo {year} {2023})}\BibitemShut {NoStop}%
\bibitem [{\citenamefont {Khalaf}\ \emph {et~al.}(2022)\citenamefont {Khalaf}, \citenamefont {Ledwith},\ and\ \citenamefont {Vishwanath}}]{Vishwanath2022PRBCharge4e}%
  \BibitemOpen
  \bibfield  {author} {\bibinfo {author} {\bibfnamefont {E.}~\bibnamefont {Khalaf}}, \bibinfo {author} {\bibfnamefont {P.}~\bibnamefont {Ledwith}},\ and\ \bibinfo {author} {\bibfnamefont {A.}~\bibnamefont {Vishwanath}},\ }\bibfield  {title} {\bibinfo {title} {Symmetry constraints on superconductivity in twisted bilayer graphene: Fractional vortices, $4e$ condensates, or nonunitary pairing},\ }\href {https://doi.org/10.1103/PhysRevB.105.224508} {\bibfield  {journal} {\bibinfo  {journal} {Phys. Rev. B}\ }\textbf {\bibinfo {volume} {105}},\ \bibinfo {pages} {224508} (\bibinfo {year} {2022})}\BibitemShut {NoStop}%
\bibitem [{\citenamefont {Zou}\ \emph {et~al.}(2025)\citenamefont {Zou}, \citenamefont {Wan},\ and\ \citenamefont {Yao}}]{Yao2025arXivcharge4e}%
  \BibitemOpen
  \bibfield  {author} {\bibinfo {author} {\bibfnamefont {X.}~\bibnamefont {Zou}}, \bibinfo {author} {\bibfnamefont {Z.-Q.}\ \bibnamefont {Wan}},\ and\ \bibinfo {author} {\bibfnamefont {H.}~\bibnamefont {Yao}},\ }\href {https://arxiv.org/abs/2510.26720} {\bibinfo {title} {Emergence of charge-$4e$ superconductivity from {2D} nematic superconductors}} (\bibinfo {year} {2025}),\ \Eprint {https://arxiv.org/abs/2510.26720} {arXiv:2510.26720 [cond-mat.supr-con]} \BibitemShut {NoStop}%
\bibitem [{\citenamefont {Lin}\ \emph {et~al.}(2025)\citenamefont {Lin}, \citenamefont {Song},\ and\ \citenamefont {Zhang}}]{Zhang2025PRB}%
  \BibitemOpen
  \bibfield  {author} {\bibinfo {author} {\bibfnamefont {T.-Y.}\ \bibnamefont {Lin}}, \bibinfo {author} {\bibfnamefont {F.-F.}\ \bibnamefont {Song}},\ and\ \bibinfo {author} {\bibfnamefont {G.-M.}\ \bibnamefont {Zhang}},\ }\bibfield  {title} {\bibinfo {title} {Theory of the charge-$6e$ condensed phase in kagome-lattice superconductors},\ }\href {https://doi.org/10.1103/PhysRevB.111.054508} {\bibfield  {journal} {\bibinfo  {journal} {Phys. Rev. B}\ }\textbf {\bibinfo {volume} {111}},\ \bibinfo {pages} {054508} (\bibinfo {year} {2025})}\BibitemShut {NoStop}%
\bibitem [{\citenamefont {Dong}\ and\ \citenamefont {Zhang}(2025)}]{Zhang2025arXiv}%
  \BibitemOpen
  \bibfield  {author} {\bibinfo {author} {\bibfnamefont {Z.-H.}\ \bibnamefont {Dong}}\ and\ \bibinfo {author} {\bibfnamefont {Y.}~\bibnamefont {Zhang}},\ }\href {https://arxiv.org/abs/2512.23801} {\bibinfo {title} {Many-electron characterizations of higher-charge superconductors}} (\bibinfo {year} {2025}),\ \Eprint {https://arxiv.org/abs/2512.23801} {arXiv:2512.23801 [cond-mat.supr-con]} \BibitemShut {NoStop}%
\bibitem [{\citenamefont {Soldini}\ \emph {et~al.}(2024)\citenamefont {Soldini}, \citenamefont {Fischer},\ and\ \citenamefont {Neupert}}]{Titus2024PRB}%
  \BibitemOpen
  \bibfield  {author} {\bibinfo {author} {\bibfnamefont {M.~O.}\ \bibnamefont {Soldini}}, \bibinfo {author} {\bibfnamefont {M.~H.}\ \bibnamefont {Fischer}},\ and\ \bibinfo {author} {\bibfnamefont {T.}~\bibnamefont {Neupert}},\ }\bibfield  {title} {\bibinfo {title} {Charge-$4e$ superconductivity in a hubbard model},\ }\href {https://doi.org/10.1103/PhysRevB.109.214509} {\bibfield  {journal} {\bibinfo  {journal} {Phys. Rev. B}\ }\textbf {\bibinfo {volume} {109}},\ \bibinfo {pages} {214509} (\bibinfo {year} {2024})}\BibitemShut {NoStop}%
\bibitem [{\citenamefont {Wu}\ \emph {et~al.}(2024)\citenamefont {Wu}, \citenamefont {Murthy},\ and\ \citenamefont {Kivelson}}]{Wu2024PRLWTe2}%
  \BibitemOpen
  \bibfield  {author} {\bibinfo {author} {\bibfnamefont {Y.-M.}\ \bibnamefont {Wu}}, \bibinfo {author} {\bibfnamefont {C.}~\bibnamefont {Murthy}},\ and\ \bibinfo {author} {\bibfnamefont {S.~A.}\ \bibnamefont {Kivelson}},\ }\bibfield  {title} {\bibinfo {title} {Possible sliding regimes in twisted bilayer {WTe}$_2$},\ }\href {https://doi.org/10.1103/PhysRevLett.133.246501} {\bibfield  {journal} {\bibinfo  {journal} {Phys. Rev. Lett.}\ }\textbf {\bibinfo {volume} {133}},\ \bibinfo {pages} {246501} (\bibinfo {year} {2024})}\BibitemShut {NoStop}%
\bibitem [{\citenamefont {Zhang}\ \emph {et~al.}(2025)\citenamefont {Zhang}, \citenamefont {Zhang},\ and\ \citenamefont {Song}}]{Song2025arXiv}%
  \BibitemOpen
  \bibfield  {author} {\bibinfo {author} {\bibfnamefont {L.}~\bibnamefont {Zhang}}, \bibinfo {author} {\bibfnamefont {Y.-H.}\ \bibnamefont {Zhang}},\ and\ \bibinfo {author} {\bibfnamefont {X.-Y.}\ \bibnamefont {Song}},\ }\href {https://arxiv.org/abs/2508.12370} {\bibinfo {title} {Charge-4$e$ anyon superconductor from doping $\text{SU}(4)_1$ chiral spin liquid}} (\bibinfo {year} {2025}),\ \Eprint {https://arxiv.org/abs/2508.12370} {arXiv:2508.12370 [cond-mat.str-el]} \BibitemShut {NoStop}%
\bibitem [{\citenamefont {Gnezdilov}\ and\ \citenamefont {Wang}(2022)}]{Wang2022PRBSYK}%
  \BibitemOpen
  \bibfield  {author} {\bibinfo {author} {\bibfnamefont {N.~V.}\ \bibnamefont {Gnezdilov}}\ and\ \bibinfo {author} {\bibfnamefont {Y.}~\bibnamefont {Wang}},\ }\bibfield  {title} {\bibinfo {title} {Solvable model for a charge-$4e$ superconductor},\ }\href {https://doi.org/10.1103/PhysRevB.106.094508} {\bibfield  {journal} {\bibinfo  {journal} {Phys. Rev. B}\ }\textbf {\bibinfo {volume} {106}},\ \bibinfo {pages} {094508} (\bibinfo {year} {2022})}\BibitemShut {NoStop}%
\bibitem [{\citenamefont {Chirolli}\ \emph {et~al.}(2024)\citenamefont {Chirolli}, \citenamefont {Braggio},\ and\ \citenamefont {Giazotto}}]{Chirolli2024PRR}%
  \BibitemOpen
  \bibfield  {author} {\bibinfo {author} {\bibfnamefont {L.}~\bibnamefont {Chirolli}}, \bibinfo {author} {\bibfnamefont {A.}~\bibnamefont {Braggio}},\ and\ \bibinfo {author} {\bibfnamefont {F.}~\bibnamefont {Giazotto}},\ }\bibfield  {title} {\bibinfo {title} {Cooper quartets in interacting hybrid superconducting systems},\ }\href {https://doi.org/10.1103/PhysRevResearch.6.033171} {\bibfield  {journal} {\bibinfo  {journal} {Phys. Rev. Res.}\ }\textbf {\bibinfo {volume} {6}},\ \bibinfo {pages} {033171} (\bibinfo {year} {2024})}\BibitemShut {NoStop}%
\bibitem [{\citenamefont {Gao}\ \emph {et~al.}(2026{\natexlab{b}})\citenamefont {Gao}, \citenamefont {Wang}, \citenamefont {Zhang},\ and\ \citenamefont {Yang}}]{Gao2026arXivcharge4e1}%
  \BibitemOpen
  \bibfield  {author} {\bibinfo {author} {\bibfnamefont {Z.-Q.}\ \bibnamefont {Gao}}, \bibinfo {author} {\bibfnamefont {Y.-Q.}\ \bibnamefont {Wang}}, \bibinfo {author} {\bibfnamefont {Y.-H.}\ \bibnamefont {Zhang}},\ and\ \bibinfo {author} {\bibfnamefont {H.}~\bibnamefont {Yang}},\ }\href {https://arxiv.org/abs/2602.03925} {\bibinfo {title} {Primary charge-4e superconductivity from doping a featureless mott insulator}} (\bibinfo {year} {2026}{\natexlab{b}}),\ \Eprint {https://arxiv.org/abs/2602.03925} {arXiv:2602.03925 [cond-mat.str-el]} \BibitemShut {NoStop}%
\bibitem [{\citenamefont {Li}\ and\ \citenamefont {Yao}(2019)}]{li2019review}%
  \BibitemOpen
  \bibfield  {author} {\bibinfo {author} {\bibfnamefont {Z.-X.}\ \bibnamefont {Li}}\ and\ \bibinfo {author} {\bibfnamefont {H.}~\bibnamefont {Yao}},\ }\bibfield  {title} {\bibinfo {title} {{Sign-problem-free Fermionic quantum Monte Carlo: developments and applications}},\ }\href {https://doi.org/10.1146/annurev-conmatphys-033117-054307} {\bibfield  {journal} {\bibinfo  {journal} {Annual Review of Condensed Matter Physics}\ }\textbf {\bibinfo {volume} {10}},\ \bibinfo {pages} {337} (\bibinfo {year} {2019})}\BibitemShut {NoStop}%
\bibitem [{\citenamefont {Assaad}\ and\ \citenamefont {Evertz}(2008)}]{AssaadReview}%
  \BibitemOpen
  \bibfield  {author} {\bibinfo {author} {\bibfnamefont {F.}~\bibnamefont {Assaad}}\ and\ \bibinfo {author} {\bibfnamefont {H.}~\bibnamefont {Evertz}},\ }\bibinfo {title} {World-line and determinantal quantum {Monte} {Carlo} methods for spins, phonons and electrons},\ in\ \href {https://doi.org/10.1007/978-3-540-74686-7_10} {\emph {\bibinfo {booktitle} {Computational Many-Particle Physics}}}\ (\bibinfo  {publisher} {Springer Berlin Heidelberg},\ \bibinfo {address} {Berlin, Heidelberg},\ \bibinfo {year} {2008})\ pp.\ \bibinfo {pages} {277--356}\BibitemShut {NoStop}%
\bibitem [{\citenamefont {Li}\ \emph {et~al.}(2017)\citenamefont {Li}, \citenamefont {Jiang}, \citenamefont {Jian},\ and\ \citenamefont {Yao}}]{Li2017NC}%
  \BibitemOpen
  \bibfield  {author} {\bibinfo {author} {\bibfnamefont {Z.-X.}\ \bibnamefont {Li}}, \bibinfo {author} {\bibfnamefont {Y.-F.}\ \bibnamefont {Jiang}}, \bibinfo {author} {\bibfnamefont {S.-K.}\ \bibnamefont {Jian}},\ and\ \bibinfo {author} {\bibfnamefont {H.}~\bibnamefont {Yao}},\ }\bibfield  {title} {\bibinfo {title} {Fermion-induced quantum critical points},\ }\href {https://doi.org/10.1038/s41467-017-00167-6} {\bibfield  {journal} {\bibinfo  {journal} {Nature Communications}\ }\textbf {\bibinfo {volume} {8}},\ \bibinfo {pages} {314} (\bibinfo {year} {2017})}\BibitemShut {NoStop}%
\bibitem [{\citenamefont {Cai}\ \emph {et~al.}(2021)\citenamefont {Cai}, \citenamefont {Li},\ and\ \citenamefont {Yao}}]{Cai2021PRLSSH}%
  \BibitemOpen
  \bibfield  {author} {\bibinfo {author} {\bibfnamefont {X.}~\bibnamefont {Cai}}, \bibinfo {author} {\bibfnamefont {Z.-X.}\ \bibnamefont {Li}},\ and\ \bibinfo {author} {\bibfnamefont {H.}~\bibnamefont {Yao}},\ }\bibfield  {title} {\bibinfo {title} {Antiferromagnetism induced by bond {Su-Schrieffer-Heeger} electron-phonon coupling: A quantum monte carlo study},\ }\href {https://doi.org/10.1103/PhysRevLett.127.247203} {\bibfield  {journal} {\bibinfo  {journal} {Phys. Rev. Lett.}\ }\textbf {\bibinfo {volume} {127}},\ \bibinfo {pages} {247203} (\bibinfo {year} {2021})}\BibitemShut {NoStop}%
\bibitem [{\citenamefont {Yu}\ \emph {et~al.}(2024)\citenamefont {Yu}, \citenamefont {Shi}, \citenamefont {Xu},\ and\ \citenamefont {Li}}]{Li2024PRLSU4}%
  \BibitemOpen
  \bibfield  {author} {\bibinfo {author} {\bibfnamefont {X.-J.}\ \bibnamefont {Yu}}, \bibinfo {author} {\bibfnamefont {S.-H.}\ \bibnamefont {Shi}}, \bibinfo {author} {\bibfnamefont {L.}~\bibnamefont {Xu}},\ and\ \bibinfo {author} {\bibfnamefont {Z.-X.}\ \bibnamefont {Li}},\ }\bibfield  {title} {\bibinfo {title} {Emergence of competing orders and possible quantum spin liquid in {{SU}(N)} fermions},\ }\href {https://doi.org/10.1103/PhysRevLett.132.036704} {\bibfield  {journal} {\bibinfo  {journal} {Phys. Rev. Lett.}\ }\textbf {\bibinfo {volume} {132}},\ \bibinfo {pages} {036704} (\bibinfo {year} {2024})}\BibitemShut {NoStop}%
\bibitem [{\citenamefont {Fradkin}\ and\ \citenamefont {Hirsch}(1983)}]{Fradkin1983PRB}%
  \BibitemOpen
  \bibfield  {author} {\bibinfo {author} {\bibfnamefont {E.}~\bibnamefont {Fradkin}}\ and\ \bibinfo {author} {\bibfnamefont {J.~E.}\ \bibnamefont {Hirsch}},\ }\bibfield  {title} {\bibinfo {title} {Phase diagram of one-dimensional electron-phonon systems. i. the {Su-Schrieffer-Heeger} model},\ }\href {https://doi.org/10.1103/PhysRevB.27.1680} {\bibfield  {journal} {\bibinfo  {journal} {Phys. Rev. B}\ }\textbf {\bibinfo {volume} {27}},\ \bibinfo {pages} {1680} (\bibinfo {year} {1983})}\BibitemShut {NoStop}%
\bibitem [{\citenamefont {Wu}\ and\ \citenamefont {Zhang}(2005)}]{Wu2005PRB}%
  \BibitemOpen
  \bibfield  {author} {\bibinfo {author} {\bibfnamefont {C.}~\bibnamefont {Wu}}\ and\ \bibinfo {author} {\bibfnamefont {S.-C.}\ \bibnamefont {Zhang}},\ }\bibfield  {title} {\bibinfo {title} {Sufficient condition for absence of the sign problem in the fermionic quantum monte carlo algorithm},\ }\href {https://doi.org/10.1103/PhysRevB.71.155115} {\bibfield  {journal} {\bibinfo  {journal} {Phys. Rev. B}\ }\textbf {\bibinfo {volume} {71}},\ \bibinfo {pages} {155115} (\bibinfo {year} {2005})}\BibitemShut {NoStop}%
\bibitem [{\citenamefont {Li}\ \emph {et~al.}(2016)\citenamefont {Li}, \citenamefont {Jiang},\ and\ \citenamefont {Yao}}]{Li2016PRL}%
  \BibitemOpen
  \bibfield  {author} {\bibinfo {author} {\bibfnamefont {Z.-X.}\ \bibnamefont {Li}}, \bibinfo {author} {\bibfnamefont {Y.-F.}\ \bibnamefont {Jiang}},\ and\ \bibinfo {author} {\bibfnamefont {H.}~\bibnamefont {Yao}},\ }\bibfield  {title} {\bibinfo {title} {Majorana-time-reversal symmetries: A fundamental principle for sign-problem-free quantum monte carlo simulations},\ }\href {https://doi.org/10.1103/PhysRevLett.117.267002} {\bibfield  {journal} {\bibinfo  {journal} {Phys. Rev. Lett.}\ }\textbf {\bibinfo {volume} {117}},\ \bibinfo {pages} {267002} (\bibinfo {year} {2016})}\BibitemShut {NoStop}%
\bibitem [{\citenamefont {Troyer}\ and\ \citenamefont {Wiese}(2005)}]{Troyer2005sign}%
  \BibitemOpen
  \bibfield  {author} {\bibinfo {author} {\bibfnamefont {M.}~\bibnamefont {Troyer}}\ and\ \bibinfo {author} {\bibfnamefont {U.-J.}\ \bibnamefont {Wiese}},\ }\bibfield  {title} {\bibinfo {title} {Computational complexity and fundamental limitations to fermionic quantum monte carlo simulations},\ }\href {https://doi.org/10.1103/PhysRevLett.94.170201} {\bibfield  {journal} {\bibinfo  {journal} {Phys. Rev. Lett.}\ }\textbf {\bibinfo {volume} {94}},\ \bibinfo {pages} {170201} (\bibinfo {year} {2005})}\BibitemShut {NoStop}%
\bibitem [{\citenamefont {Sorella}\ \emph {et~al.}(1989)\citenamefont {Sorella}, \citenamefont {Baroni}, \citenamefont {Car},\ and\ \citenamefont {Parrinello}}]{Sorella1989EPL}%
  \BibitemOpen
  \bibfield  {author} {\bibinfo {author} {\bibfnamefont {S.}~\bibnamefont {Sorella}}, \bibinfo {author} {\bibfnamefont {S.}~\bibnamefont {Baroni}}, \bibinfo {author} {\bibfnamefont {R.}~\bibnamefont {Car}},\ and\ \bibinfo {author} {\bibfnamefont {M.}~\bibnamefont {Parrinello}},\ }\bibfield  {title} {\bibinfo {title} {A novel technique for the simulation of interacting fermion systems},\ }\href {https://doi.org/10.1209/0295-5075/8/7/014} {\bibfield  {journal} {\bibinfo  {journal} {Europhysics Letters}\ }\textbf {\bibinfo {volume} {8}},\ \bibinfo {pages} {663} (\bibinfo {year} {1989})}\BibitemShut {NoStop}%
\bibitem [{\citenamefont {Blankenbecler}\ \emph {et~al.}(1981)\citenamefont {Blankenbecler}, \citenamefont {Scalapino},\ and\ \citenamefont {Sugar}}]{BSS}%
  \BibitemOpen
  \bibfield  {author} {\bibinfo {author} {\bibfnamefont {R.}~\bibnamefont {Blankenbecler}}, \bibinfo {author} {\bibfnamefont {D.~J.}\ \bibnamefont {Scalapino}},\ and\ \bibinfo {author} {\bibfnamefont {R.~L.}\ \bibnamefont {Sugar}},\ }\bibfield  {title} {\bibinfo {title} {Monte carlo calculations of coupled boson-fermion systems. i},\ }\href {https://doi.org/10.1103/PhysRevD.24.2278} {\bibfield  {journal} {\bibinfo  {journal} {Phys. Rev. D}\ }\textbf {\bibinfo {volume} {24}},\ \bibinfo {pages} {2278} (\bibinfo {year} {1981})}\BibitemShut {NoStop}%
\bibitem [{\citenamefont {Scalapino}\ \emph {et~al.}(1993)\citenamefont {Scalapino}, \citenamefont {White},\ and\ \citenamefont {Zhang}}]{Zhang1993PRB}%
  \BibitemOpen
  \bibfield  {author} {\bibinfo {author} {\bibfnamefont {D.~J.}\ \bibnamefont {Scalapino}}, \bibinfo {author} {\bibfnamefont {S.~R.}\ \bibnamefont {White}},\ and\ \bibinfo {author} {\bibfnamefont {S.}~\bibnamefont {Zhang}},\ }\bibfield  {title} {\bibinfo {title} {Insulator, metal, or superconductor: The criteria},\ }\href {https://doi.org/10.1103/PhysRevB.47.7995} {\bibfield  {journal} {\bibinfo  {journal} {Phys. Rev. B}\ }\textbf {\bibinfo {volume} {47}},\ \bibinfo {pages} {7995} (\bibinfo {year} {1993})}\BibitemShut {NoStop}%
\bibitem [{\citenamefont {Anderson}(1987)}]{Anderson1987Science}%
  \BibitemOpen
  \bibfield  {author} {\bibinfo {author} {\bibfnamefont {P.~W.}\ \bibnamefont {Anderson}},\ }\bibfield  {title} {\bibinfo {title} {The resonating valence bond state in {La$_2$CuO$_4$} and superconductivity},\ }\href {https://doi.org/10.1126/science.235.4793.1196} {\bibfield  {journal} {\bibinfo  {journal} {Science}\ }\textbf {\bibinfo {volume} {235}},\ \bibinfo {pages} {1196} (\bibinfo {year} {1987})}\BibitemShut {NoStop}%
\bibitem [{\citenamefont {Berezinskii}(1972)}]{BKT1972}%
  \BibitemOpen
  \bibfield  {author} {\bibinfo {author} {\bibfnamefont {V.~L.}\ \bibnamefont {Berezinskii}},\ }\bibfield  {title} {\bibinfo {title} {Destruction of long-range order in one-dimensional and two-dimensional systems possessing a continuous symmetry group. ii. quantum systems},\ }\href {https://www.jetp.ras.ru/cgi-bin/e/index/e/34/3/p610?a=list} {\bibfield  {journal} {\bibinfo  {journal} {Sov. Phys. JETP}\ }\textbf {\bibinfo {volume} {34}},\ \bibinfo {pages} {610} (\bibinfo {year} {1972})}\BibitemShut {NoStop}%
\bibitem [{\citenamefont {Kosterlitz}\ and\ \citenamefont {Thouless}(1973)}]{KT1973}%
  \BibitemOpen
  \bibfield  {author} {\bibinfo {author} {\bibfnamefont {J.~M.}\ \bibnamefont {Kosterlitz}}\ and\ \bibinfo {author} {\bibfnamefont {D.~J.}\ \bibnamefont {Thouless}},\ }\bibfield  {title} {\bibinfo {title} {Ordering, metastability and phase transitions in two-dimensional systems},\ }\href {https://doi.org/10.1088/0022-3719/6/7/010} {\bibfield  {journal} {\bibinfo  {journal} {Journal of Physics C: Solid State Physics}\ }\textbf {\bibinfo {volume} {6}},\ \bibinfo {pages} {1181} (\bibinfo {year} {1973})}\BibitemShut {NoStop}%
\bibitem [{\citenamefont {Kosterlitz}(1974)}]{KT1974}%
  \BibitemOpen
  \bibfield  {author} {\bibinfo {author} {\bibfnamefont {J.~M.}\ \bibnamefont {Kosterlitz}},\ }\bibfield  {title} {\bibinfo {title} {The critical properties of the two-dimensional {XY} model},\ }\href {https://doi.org/10.1088/0022-3719/7/6/005} {\bibfield  {journal} {\bibinfo  {journal} {Journal of Physics C: Solid State Physics}\ }\textbf {\bibinfo {volume} {7}},\ \bibinfo {pages} {1046} (\bibinfo {year} {1974})}\BibitemShut {NoStop}%
\bibitem [{\citenamefont {Lee}\ and\ \citenamefont {Grinstein}(1985)}]{Lee1985PRB}%
  \BibitemOpen
  \bibfield  {author} {\bibinfo {author} {\bibfnamefont {D.~H.}\ \bibnamefont {Lee}}\ and\ \bibinfo {author} {\bibfnamefont {G.}~\bibnamefont {Grinstein}},\ }\bibfield  {title} {\bibinfo {title} {Strings in two-dimensional classical {XY} models},\ }\href {https://doi.org/10.1103/PhysRevLett.55.541} {\bibfield  {journal} {\bibinfo  {journal} {Phys. Rev. Lett.}\ }\textbf {\bibinfo {volume} {55}},\ \bibinfo {pages} {541} (\bibinfo {year} {1985})}\BibitemShut {NoStop}%
\bibitem [{\citenamefont {Cai}\ \emph {et~al.}(2025)\citenamefont {Cai}, \citenamefont {Li},\ and\ \citenamefont {Yao}}]{Cai2025PRB}%
  \BibitemOpen
  \bibfield  {author} {\bibinfo {author} {\bibfnamefont {X.}~\bibnamefont {Cai}}, \bibinfo {author} {\bibfnamefont {Z.-X.}\ \bibnamefont {Li}},\ and\ \bibinfo {author} {\bibfnamefont {H.}~\bibnamefont {Yao}},\ }\bibfield  {title} {\bibinfo {title} {High-temperature superconductivity induced by the {Su-Schrieffer-Heeger} electron-phonon coupling},\ }\href {https://doi.org/10.1103/rhss-d52m} {\bibfield  {journal} {\bibinfo  {journal} {Phys. Rev. B}\ }\textbf {\bibinfo {volume} {112}},\ \bibinfo {pages} {144517} (\bibinfo {year} {2025})}\BibitemShut {NoStop}%
\bibitem [{\citenamefont {Zhang}\ \emph {et~al.}(2023)\citenamefont {Zhang}, \citenamefont {Sous}, \citenamefont {Reichman}, \citenamefont {Berciu}, \citenamefont {Millis}, \citenamefont {Prokof'ev},\ and\ \citenamefont {Svistunov}}]{Zhang2023PRXBipolaron}%
  \BibitemOpen
  \bibfield  {author} {\bibinfo {author} {\bibfnamefont {C.}~\bibnamefont {Zhang}}, \bibinfo {author} {\bibfnamefont {J.}~\bibnamefont {Sous}}, \bibinfo {author} {\bibfnamefont {D.~R.}\ \bibnamefont {Reichman}}, \bibinfo {author} {\bibfnamefont {M.}~\bibnamefont {Berciu}}, \bibinfo {author} {\bibfnamefont {A.~J.}\ \bibnamefont {Millis}}, \bibinfo {author} {\bibfnamefont {N.~V.}\ \bibnamefont {Prokof'ev}},\ and\ \bibinfo {author} {\bibfnamefont {B.~V.}\ \bibnamefont {Svistunov}},\ }\bibfield  {title} {\bibinfo {title} {Bipolaronic high-temperature superconductivity},\ }\href {https://doi.org/10.1103/PhysRevX.13.011010} {\bibfield  {journal} {\bibinfo  {journal} {Phys. Rev. X}\ }\textbf {\bibinfo {volume} {13}},\ \bibinfo {pages} {011010} (\bibinfo {year} {2023})}\BibitemShut {NoStop}%
\bibitem [{\citenamefont {Sandvik}(1998)}]{sandvik1998SAC}%
  \BibitemOpen
  \bibfield  {author} {\bibinfo {author} {\bibfnamefont {A.~W.}\ \bibnamefont {Sandvik}},\ }\bibfield  {title} {\bibinfo {title} {Stochastic method for analytic continuation of quantum monte carlo data},\ }\href {https://doi.org/10.1103/PhysRevB.57.10287} {\bibfield  {journal} {\bibinfo  {journal} {Phys. Rev. B}\ }\textbf {\bibinfo {volume} {57}},\ \bibinfo {pages} {10287} (\bibinfo {year} {1998})}\BibitemShut {NoStop}%
\bibitem [{\citenamefont {Chen}\ \emph {et~al.}(2022)\citenamefont {Chen}, \citenamefont {Hashimoto}, \citenamefont {He}, \citenamefont {Song}, \citenamefont {He}, \citenamefont {Li}, \citenamefont {Ishida}, \citenamefont {Eisaki}, \citenamefont {Zaanen}, \citenamefont {Devereaux}, \citenamefont {Lee}, \citenamefont {Lu},\ and\ \citenamefont {Shen}}]{ZXShen2022Nature}%
  \BibitemOpen
  \bibfield  {author} {\bibinfo {author} {\bibfnamefont {S.-D.}\ \bibnamefont {Chen}}, \bibinfo {author} {\bibfnamefont {M.}~\bibnamefont {Hashimoto}}, \bibinfo {author} {\bibfnamefont {Y.}~\bibnamefont {He}}, \bibinfo {author} {\bibfnamefont {D.}~\bibnamefont {Song}}, \bibinfo {author} {\bibfnamefont {J.-F.}\ \bibnamefont {He}}, \bibinfo {author} {\bibfnamefont {Y.-F.}\ \bibnamefont {Li}}, \bibinfo {author} {\bibfnamefont {S.}~\bibnamefont {Ishida}}, \bibinfo {author} {\bibfnamefont {H.}~\bibnamefont {Eisaki}}, \bibinfo {author} {\bibfnamefont {J.}~\bibnamefont {Zaanen}}, \bibinfo {author} {\bibfnamefont {T.~P.}\ \bibnamefont {Devereaux}}, \bibinfo {author} {\bibfnamefont {D.-H.}\ \bibnamefont {Lee}}, \bibinfo {author} {\bibfnamefont {D.-H.}\ \bibnamefont {Lu}},\ and\ \bibinfo {author} {\bibfnamefont {Z.-X.}\ \bibnamefont {Shen}},\ }\bibfield  {title} {\bibinfo {title} {Unconventional spectral signature of {$T_c$} in a pure d-wave superconductor},\ }\href {https://doi.org/10.1038/s41586-021-04251-2}
  {\bibfield  {journal} {\bibinfo  {journal} {Nature}\ }\textbf {\bibinfo {volume} {601}},\ \bibinfo {pages} {562} (\bibinfo {year} {2022})}\BibitemShut {NoStop}%
\bibitem [{\citenamefont {He}\ \emph {et~al.}(2021)\citenamefont {He}, \citenamefont {Chen}, \citenamefont {Li}, \citenamefont {Zhao}, \citenamefont {Song}, \citenamefont {Yoshida}, \citenamefont {Eisaki}, \citenamefont {Wu}, \citenamefont {Chen}, \citenamefont {Lu}, \citenamefont {Meingast}, \citenamefont {Devereaux}, \citenamefont {Birgeneau}, \citenamefont {Hashimoto}, \citenamefont {Lee},\ and\ \citenamefont {Shen}}]{He2021PRXSCfluctuation}%
  \BibitemOpen
  \bibfield  {author} {\bibinfo {author} {\bibfnamefont {Y.}~\bibnamefont {He}}, \bibinfo {author} {\bibfnamefont {S.-D.}\ \bibnamefont {Chen}}, \bibinfo {author} {\bibfnamefont {Z.-X.}\ \bibnamefont {Li}}, \bibinfo {author} {\bibfnamefont {D.}~\bibnamefont {Zhao}}, \bibinfo {author} {\bibfnamefont {D.}~\bibnamefont {Song}}, \bibinfo {author} {\bibfnamefont {Y.}~\bibnamefont {Yoshida}}, \bibinfo {author} {\bibfnamefont {H.}~\bibnamefont {Eisaki}}, \bibinfo {author} {\bibfnamefont {T.}~\bibnamefont {Wu}}, \bibinfo {author} {\bibfnamefont {X.-H.}\ \bibnamefont {Chen}}, \bibinfo {author} {\bibfnamefont {D.-H.}\ \bibnamefont {Lu}}, \bibinfo {author} {\bibfnamefont {C.}~\bibnamefont {Meingast}}, \bibinfo {author} {\bibfnamefont {T.~P.}\ \bibnamefont {Devereaux}}, \bibinfo {author} {\bibfnamefont {R.~J.}\ \bibnamefont {Birgeneau}}, \bibinfo {author} {\bibfnamefont {M.}~\bibnamefont {Hashimoto}}, \bibinfo {author} {\bibfnamefont {D.-H.}\ \bibnamefont {Lee}},\ and\ \bibinfo {author} {\bibfnamefont {Z.-X.}\
  \bibnamefont {Shen}},\ }\bibfield  {title} {\bibinfo {title} {Superconducting fluctuations in overdoped {Bi$_2$Sr$_2$CaCu$_2$O$_{8+\delta}$}},\ }\href {https://doi.org/10.1103/PhysRevX.11.031068} {\bibfield  {journal} {\bibinfo  {journal} {Phys. Rev. X}\ }\textbf {\bibinfo {volume} {11}},\ \bibinfo {pages} {031068} (\bibinfo {year} {2021})}\BibitemShut {NoStop}%
\bibitem [{\citenamefont {Emery}\ and\ \citenamefont {Kivelson}(1995)}]{Kivelson1995Nature}%
  \BibitemOpen
  \bibfield  {author} {\bibinfo {author} {\bibfnamefont {V.~J.}\ \bibnamefont {Emery}}\ and\ \bibinfo {author} {\bibfnamefont {S.~A.}\ \bibnamefont {Kivelson}},\ }\bibfield  {title} {\bibinfo {title} {Importance of phase fluctuations in superconductors with small superfluid density},\ }\href {https://doi.org/10.1038/374434a0} {\bibfield  {journal} {\bibinfo  {journal} {Nature}\ }\textbf {\bibinfo {volume} {374}},\ \bibinfo {pages} {434} (\bibinfo {year} {1995})}\BibitemShut {NoStop}%
\bibitem [{\citenamefont {Taie}\ \emph {et~al.}(2010)\citenamefont {Taie}, \citenamefont {Takasu}, \citenamefont {Sugawa}, \citenamefont {Yamazaki}, \citenamefont {Tsujimoto}, \citenamefont {Murakami},\ and\ \citenamefont {Takahashi}}]{Yoshiro2010PRLSUN}%
  \BibitemOpen
  \bibfield  {author} {\bibinfo {author} {\bibfnamefont {S.}~\bibnamefont {Taie}}, \bibinfo {author} {\bibfnamefont {Y.}~\bibnamefont {Takasu}}, \bibinfo {author} {\bibfnamefont {S.}~\bibnamefont {Sugawa}}, \bibinfo {author} {\bibfnamefont {R.}~\bibnamefont {Yamazaki}}, \bibinfo {author} {\bibfnamefont {T.}~\bibnamefont {Tsujimoto}}, \bibinfo {author} {\bibfnamefont {R.}~\bibnamefont {Murakami}},\ and\ \bibinfo {author} {\bibfnamefont {Y.}~\bibnamefont {Takahashi}},\ }\bibfield  {title} {\bibinfo {title} {Realization of a $\mathrm{SU}(2)\ifmmode\times\else\texttimes\fi{}\mathrm{SU}(6)$ system of fermions in a cold atomic gas},\ }\href {https://doi.org/10.1103/PhysRevLett.105.190401} {\bibfield  {journal} {\bibinfo  {journal} {Phys. Rev. Lett.}\ }\textbf {\bibinfo {volume} {105}},\ \bibinfo {pages} {190401} (\bibinfo {year} {2010})}\BibitemShut {NoStop}%
\bibitem [{\citenamefont {Krauser}\ \emph {et~al.}(2012)\citenamefont {Krauser}, \citenamefont {Heinze}, \citenamefont {Fl{\"a}schner}, \citenamefont {G{\"o}tze}, \citenamefont {J{\"u}rgensen}, \citenamefont {L{\"u}hmann}, \citenamefont {Becker},\ and\ \citenamefont {Sengstock}}]{Sengstock2012NPSUN}%
  \BibitemOpen
  \bibfield  {author} {\bibinfo {author} {\bibfnamefont {J.~S.}\ \bibnamefont {Krauser}}, \bibinfo {author} {\bibfnamefont {J.}~\bibnamefont {Heinze}}, \bibinfo {author} {\bibfnamefont {N.}~\bibnamefont {Fl{\"a}schner}}, \bibinfo {author} {\bibfnamefont {S.}~\bibnamefont {G{\"o}tze}}, \bibinfo {author} {\bibfnamefont {O.}~\bibnamefont {J{\"u}rgensen}}, \bibinfo {author} {\bibfnamefont {D.-S.}\ \bibnamefont {L{\"u}hmann}}, \bibinfo {author} {\bibfnamefont {C.}~\bibnamefont {Becker}},\ and\ \bibinfo {author} {\bibfnamefont {K.}~\bibnamefont {Sengstock}},\ }\bibfield  {title} {\bibinfo {title} {Coherent multi-flavour spin dynamics in a fermionic quantum gas},\ }\href {https://doi.org/10.1038/nphys2409} {\bibfield  {journal} {\bibinfo  {journal} {Nature Physics}\ }\textbf {\bibinfo {volume} {8}},\ \bibinfo {pages} {813} (\bibinfo {year} {2012})}\BibitemShut {NoStop}%
\bibitem [{\citenamefont {Pagano}\ \emph {et~al.}(2014)\citenamefont {Pagano}, \citenamefont {Mancini}, \citenamefont {Cappellini}, \citenamefont {Lombardi}, \citenamefont {Sch{\"a}fer}, \citenamefont {Hu}, \citenamefont {Liu}, \citenamefont {Catani}, \citenamefont {Sias}, \citenamefont {Inguscio},\ and\ \citenamefont {Fallani}}]{Fallani2014NPSUN}%
  \BibitemOpen
  \bibfield  {author} {\bibinfo {author} {\bibfnamefont {G.}~\bibnamefont {Pagano}}, \bibinfo {author} {\bibfnamefont {M.}~\bibnamefont {Mancini}}, \bibinfo {author} {\bibfnamefont {G.}~\bibnamefont {Cappellini}}, \bibinfo {author} {\bibfnamefont {P.}~\bibnamefont {Lombardi}}, \bibinfo {author} {\bibfnamefont {F.}~\bibnamefont {Sch{\"a}fer}}, \bibinfo {author} {\bibfnamefont {H.}~\bibnamefont {Hu}}, \bibinfo {author} {\bibfnamefont {X.-J.}\ \bibnamefont {Liu}}, \bibinfo {author} {\bibfnamefont {J.}~\bibnamefont {Catani}}, \bibinfo {author} {\bibfnamefont {C.}~\bibnamefont {Sias}}, \bibinfo {author} {\bibfnamefont {M.}~\bibnamefont {Inguscio}},\ and\ \bibinfo {author} {\bibfnamefont {L.}~\bibnamefont {Fallani}},\ }\bibfield  {title} {\bibinfo {title} {A one-dimensional liquid of fermions with tunable spin},\ }\href {https://doi.org/10.1038/nphys2878} {\bibfield  {journal} {\bibinfo  {journal} {Nature Physics}\ }\textbf {\bibinfo {volume} {10}},\ \bibinfo {pages} {198} (\bibinfo {year} {2014})}\BibitemShut
  {NoStop}%
\bibitem [{\citenamefont {Taie}\ \emph {et~al.}(2012)\citenamefont {Taie}, \citenamefont {Yamazaki}, \citenamefont {Sugawa},\ and\ \citenamefont {Takahashi}}]{Yoshiro2012NPSU6}%
  \BibitemOpen
  \bibfield  {author} {\bibinfo {author} {\bibfnamefont {S.}~\bibnamefont {Taie}}, \bibinfo {author} {\bibfnamefont {R.}~\bibnamefont {Yamazaki}}, \bibinfo {author} {\bibfnamefont {S.}~\bibnamefont {Sugawa}},\ and\ \bibinfo {author} {\bibfnamefont {Y.}~\bibnamefont {Takahashi}},\ }\bibfield  {title} {\bibinfo {title} {An {SU(6)} {Mott} insulator of an atomic fermi gas realized by large-spin {Pomeranchuk} cooling},\ }\href {https://doi.org/10.1038/nphys2430} {\bibfield  {journal} {\bibinfo  {journal} {Nature Physics}\ }\textbf {\bibinfo {volume} {8}},\ \bibinfo {pages} {825} (\bibinfo {year} {2012})}\BibitemShut {NoStop}%
\bibitem [{\citenamefont {Xu}\ \emph {et~al.}(2018)\citenamefont {Xu}, \citenamefont {Barreiro}, \citenamefont {Wang},\ and\ \citenamefont {Wu}}]{Wu2018PRLSUN}%
  \BibitemOpen
  \bibfield  {author} {\bibinfo {author} {\bibfnamefont {S.}~\bibnamefont {Xu}}, \bibinfo {author} {\bibfnamefont {J.~T.}\ \bibnamefont {Barreiro}}, \bibinfo {author} {\bibfnamefont {Y.}~\bibnamefont {Wang}},\ and\ \bibinfo {author} {\bibfnamefont {C.}~\bibnamefont {Wu}},\ }\bibfield  {title} {\bibinfo {title} {Interaction effects with varying {$N$} in {SU(N)} symmetric fermion lattice systems},\ }\href {https://doi.org/10.1103/PhysRevLett.121.167205} {\bibfield  {journal} {\bibinfo  {journal} {Phys. Rev. Lett.}\ }\textbf {\bibinfo {volume} {121}},\ \bibinfo {pages} {167205} (\bibinfo {year} {2018})}\BibitemShut {NoStop}%
\bibitem [{\citenamefont {Chen}\ and\ \citenamefont {Wu}(2024)}]{Wu2024npjQMReview}%
  \BibitemOpen
  \bibfield  {author} {\bibinfo {author} {\bibfnamefont {G.~V.}\ \bibnamefont {Chen}}\ and\ \bibinfo {author} {\bibfnamefont {C.}~\bibnamefont {Wu}},\ }\bibfield  {title} {\bibinfo {title} {Multiflavor mott insulators in quantum materials and ultracold atoms},\ }\href {https://doi.org/10.1038/s41535-023-00614-2} {\bibfield  {journal} {\bibinfo  {journal} {npj Quantum Materials}\ }\textbf {\bibinfo {volume} {9}},\ \bibinfo {pages} {1} (\bibinfo {year} {2024})}\BibitemShut {NoStop}%
\bibitem [{\citenamefont {Shibauchi}\ \emph {et~al.}(2014)\citenamefont {Shibauchi}, \citenamefont {Carrington},\ and\ \citenamefont {Matsuda}}]{Matsuda2014Review}%
  \BibitemOpen
  \bibfield  {author} {\bibinfo {author} {\bibfnamefont {T.}~\bibnamefont {Shibauchi}}, \bibinfo {author} {\bibfnamefont {A.}~\bibnamefont {Carrington}},\ and\ \bibinfo {author} {\bibfnamefont {Y.}~\bibnamefont {Matsuda}},\ }\bibfield  {title} {\bibinfo {title} {A quantum critical point lying beneath the superconducting dome in iron pnictides},\ }\href {https://doi.org/https://doi.org/10.1146/annurev-conmatphys-031113-133921} {\bibfield  {journal} {\bibinfo  {journal} {Annual Review of Condensed Matter Physics}\ }\textbf {\bibinfo {volume} {5}},\ \bibinfo {pages} {113} (\bibinfo {year} {2014})}\BibitemShut {NoStop}%
\bibitem [{\citenamefont {Xu}\ and\ \citenamefont {Balents}(2018)}]{Xu2018PRL}%
  \BibitemOpen
  \bibfield  {author} {\bibinfo {author} {\bibfnamefont {C.}~\bibnamefont {Xu}}\ and\ \bibinfo {author} {\bibfnamefont {L.}~\bibnamefont {Balents}},\ }\bibfield  {title} {\bibinfo {title} {Topological superconductivity in twisted multilayer graphene},\ }\href {https://doi.org/10.1103/PhysRevLett.121.087001} {\bibfield  {journal} {\bibinfo  {journal} {Phys. Rev. Lett.}\ }\textbf {\bibinfo {volume} {121}},\ \bibinfo {pages} {087001} (\bibinfo {year} {2018})}\BibitemShut {NoStop}%
\bibitem [{\citenamefont {Zhang}\ \emph {et~al.}(2021)\citenamefont {Zhang}, \citenamefont {Sheng},\ and\ \citenamefont {Vishwanath}}]{Zhang2021PRLSU4}%
  \BibitemOpen
  \bibfield  {author} {\bibinfo {author} {\bibfnamefont {Y.-H.}\ \bibnamefont {Zhang}}, \bibinfo {author} {\bibfnamefont {D.~N.}\ \bibnamefont {Sheng}},\ and\ \bibinfo {author} {\bibfnamefont {A.}~\bibnamefont {Vishwanath}},\ }\bibfield  {title} {\bibinfo {title} {{SU(4)} chiral spin liquid, exciton supersolid, and electric detection in moir\'e bilayers},\ }\href {https://doi.org/10.1103/PhysRevLett.127.247701} {\bibfield  {journal} {\bibinfo  {journal} {Phys. Rev. Lett.}\ }\textbf {\bibinfo {volume} {127}},\ \bibinfo {pages} {247701} (\bibinfo {year} {2021})}\BibitemShut {NoStop}%
\bibitem [{\citenamefont {Parthenios}\ and\ \citenamefont {Classen}(2023)}]{Classen2023PRB}%
  \BibitemOpen
  \bibfield  {author} {\bibinfo {author} {\bibfnamefont {N.}~\bibnamefont {Parthenios}}\ and\ \bibinfo {author} {\bibfnamefont {L.}~\bibnamefont {Classen}},\ }\bibfield  {title} {\bibinfo {title} {Twisted bilayer graphene at charge neutrality: Competing orders of {$\mathrm{SU}(4)$} {Dirac} fermions},\ }\href {https://doi.org/10.1103/PhysRevB.108.235120} {\bibfield  {journal} {\bibinfo  {journal} {Phys. Rev. B}\ }\textbf {\bibinfo {volume} {108}},\ \bibinfo {pages} {235120} (\bibinfo {year} {2023})}\BibitemShut {NoStop}%
\bibitem [{\citenamefont {Wan}\ \emph {et~al.}(2026)\citenamefont {Wan}, \citenamefont {Jiang}, \citenamefont {Zou}, \citenamefont {Zhang},\ and\ \citenamefont {Jian}}]{Jiancharge4e}%
  \BibitemOpen
  \bibfield  {author} {\bibinfo {author} {\bibfnamefont {Z.-Q.}\ \bibnamefont {Wan}}, \bibinfo {author} {\bibfnamefont {H.}~\bibnamefont {Jiang}}, \bibinfo {author} {\bibfnamefont {X.}~\bibnamefont {Zou}}, \bibinfo {author} {\bibfnamefont {S.}~\bibnamefont {Zhang}},\ and\ \bibinfo {author} {\bibfnamefont {S.-K.}\ \bibnamefont {Jian}},\ }\bibfield  {title} {\bibinfo {title} {Quantum charge-$4e$ superconductivity and deconfined pseudocriticality in the attractive {SU}(4) hubbard model}} (\bibinfo {year} {2026})\BibitemShut {NoStop}%
\bibitem [{\citenamefont {Miles}\ \emph {et~al.}(2022)\citenamefont {Miles}, \citenamefont {Cohen-Stead}, \citenamefont {Bradley}, \citenamefont {Johnston}, \citenamefont {Scalettar},\ and\ \citenamefont {Barros}}]{mu_tune}%
  \BibitemOpen
  \bibfield  {author} {\bibinfo {author} {\bibfnamefont {C.}~\bibnamefont {Miles}}, \bibinfo {author} {\bibfnamefont {B.}~\bibnamefont {Cohen-Stead}}, \bibinfo {author} {\bibfnamefont {O.}~\bibnamefont {Bradley}}, \bibinfo {author} {\bibfnamefont {S.}~\bibnamefont {Johnston}}, \bibinfo {author} {\bibfnamefont {R.}~\bibnamefont {Scalettar}},\ and\ \bibinfo {author} {\bibfnamefont {K.}~\bibnamefont {Barros}},\ }\bibfield  {title} {\bibinfo {title} {Dynamical tuning of the chemical potential to achieve a target particle number in grand canonical monte carlo simulations},\ }\href {https://doi.org/10.1103/PhysRevE.105.045311} {\bibfield  {journal} {\bibinfo  {journal} {Phys. Rev. E}\ }\textbf {\bibinfo {volume} {105}},\ \bibinfo {pages} {045311} (\bibinfo {year} {2022})}\BibitemShut {NoStop}%
\bibitem [{\citenamefont {Shao}\ and\ \citenamefont {Sandvik}(2023)}]{shao2023SAC}%
  \BibitemOpen
  \bibfield  {author} {\bibinfo {author} {\bibfnamefont {H.}~\bibnamefont {Shao}}\ and\ \bibinfo {author} {\bibfnamefont {A.~W.}\ \bibnamefont {Sandvik}},\ }\bibfield  {title} {\bibinfo {title} {Progress on stochastic analytic continuation of quantum monte carlo data},\ }\href {https://doi.org/10.1016/j.physrep.2022.11.002} {\bibfield  {journal} {\bibinfo  {journal} {Phys. Rep.}\ }\textbf {\bibinfo {volume} {1003}},\ \bibinfo {pages} {1} (\bibinfo {year} {2023})}\BibitemShut {NoStop}%
\end{thebibliography}%

\clearpage
\onecolumngrid

\begin{center}
\textbf{\large Supplemental Material for\\[4pt]
``High-temperature charge-$4e$ superconductivity in SU(4) interacting fermions''}
\end{center}

\setcounter{equation}{0}
\renewcommand{\theequation}{S\arabic{equation}}

\setcounter{figure}{0}
\renewcommand{\thefigure}{S\arabic{figure}}

\setcounter{table}{0}
\renewcommand{\thetable}{S\arabic{table}}


This Supplementary Material provides additional technical details, numerical analyses, and supporting results for the main text. It is organized into six sections.

Section~I presents the details of the quantum Monte Carlo methods employed in this work, including both the projector quantum Monte Carlo (PQMC) and the finite-temperature quantum Monte Carlo (FTQMC) algorithms. 

Section II presents a finite-size scaling analysis of the charge-$2e$ and charge-$4e$ pairing order parameters. By extrapolating to the thermodynamic limit, this analysis confirms the suppression of charge-$2e$ order and the persistence of charge-$4e$ order in the strong-coupling regime.

In Section~III, we derive the explicit form of the current operator for the model, which is required for the evaluation of superfluid stiffness. 


Section~IV considers the case in which charge-$2e$ condensation occurs and analyzes the topology of the order-parameter manifold. We show that the broken $\mathrm{SU}(4)$ symmetry supports thermally excited half-vortex defects, which in turn imply a finite-temperature charge-$4e$ phase, together with a Berezinskii-Kosterlitz-Thouless (BKT) transition characterized by a universal stiffness jump at a critical temperature determined by $\rho_s(T)=8T/\pi$.

Section V provides direct numerical evidence for this BKT transition by examining the temperature dependence of the superfluid stiffness. We also present the single-particle spectral function, revealing a robust pseudogap that persists above the critical temperature. We also show that the finite-temperature single-particle spectral function retains a robust pseudogap.

Finally, Section~VI provides a brief overview of the stochastic analytic continuation method used to extract the single-particle spectral functions from imaginary-time data.

\section{Section~I.~Determinantal Quantum Monte Carlo Methods}
This section details the determinantal quantum Monte Carlo (DQMC) framework used in our study. For the ground-state properties, we employ the PQMC approach. This allows us to map out the zero-temperature phase diagram by computing equal-time pairing correlations for both charge-2e and charge-4e channels. Additionally, we calculate the zero-temperature superfluid stiffness from current-current correlations to confirm the presence of a true superconducting state. To study the system at finite temperatures, we switch to the FTQMC algorithm. With FTQMC, we calculate the temperature-dependent superfluid stiffness to precisely identify and characterize the half-vortex-driven BKT transition. Finally, we compute the imaginary-time single-particle Green's function, which serves as the input for the stochastic analytic continuation method used to obtain the real-frequency spectral function.

For an observable $\hat{O}$, its ground-state expectation value is obtained by projecting a trial wave function $|\psi_T\rangle$ along the imaginary-time direction:
\begin{equation}
\langle \hat{O} \rangle = \frac{\langle \psi_T | e^{-\Theta \hat{H}} \, \hat{O} \, e^{-\Theta \hat{H}} | \psi_T \rangle}{\langle \psi_T | e^{-2\Theta \hat{H}} | \psi_T \rangle},
\label{eqp}
\end{equation}
where the trial wave function $|\psi_T\rangle$ is assumed to have a nonzero overlap with the true ground state, and the projection time $\Theta$ is chosen sufficiently large to ensure convergence to the ground state. 
At finite temperature, the expectation value of $\hat{O}$ is obtained from the grand-canonical ensemble,
\begin{equation}
\langle \hat{O} \rangle = \frac{\mathrm{Tr}\left( e^{-\beta (\hat{H} - \mu \hat{N})} \hat{O} \right)}{\mathrm{Tr}\left( e^{-\beta (\hat{H} - \mu \hat{N})} \right)},
\label{eqft}
\end{equation}
where $\beta = 1/T$ is the inverse temperature and $\mu$ is the chemical potential used to tune the particle density. Here the Hamiltonian can be decomposed as
\begin{equation}
\hat{H} = \hat{H}_t + \hat{H}_{\mathrm{int}},
\end{equation}
with $\hat{H}_t = -t \sum_{\langle ij\rangle} (\hat{c}^{\dagger}_{i\sigma} \hat{c}_{j\sigma} + \mathrm{h.c.})$ describing the single-body tight-binding part and $\hat{H}_{\mathrm{int}} = - \frac{J}{2N} \sum_{\langle ij \rangle} (\hat{c}^{\dagger}_{i\sigma} \hat{c}_{j\sigma} + \mathrm{h.c.})^2 $ accounting for the two-body interaction term. In the following, we separately describe how to evaluate Eqs.~\eqref{eqp} using PQMC and \eqref{eqft} using FTQMC. 

Within PQMC, we first discretize the imaginary-time propagation into $m$ small time slices and then apply the Trotter--Suzuki decomposition to separate $H_t$  from $H_{\mathrm{int}}$,
\begin{equation}
  \langle \psi_T | e^{-2\Theta \hat{H}} | \psi_T \rangle = \langle \psi_T | (e^{-\Delta_{\tau} \hat{H}})^m | \psi_T \rangle = \langle \psi_T | (e^{-\Delta_{\tau} H_{\mathrm{int}}} e^{-\Delta_{\tau} \hat{H_t}})^m | \psi_T \rangle  + \mathcal{O}(\Delta_\tau^2)
\end{equation}
where $m\Delta_{\tau} = 2\Theta$.  
To facilitate the numerical implementation, we employ a checkerboard decomposition of the interaction term. The Hamiltonian $\hat{H}_{\mathrm{int}}$ is partitioned into $N_c=4$ groups,
\begin{equation}
\hat{H}_{\mathrm{int}} = \sum_{c=1}^{N_c} \hat{H}_c = \sum_{c=1}^{N_c} (\sum_{b \in c}\hat{H}_{c,b}),
\end{equation}
where each $\hat{H}_c$ consists of a set of bond interaction terms $\hat{H}_{c,b}$  arranged such that no two bonds within the same group share a common site. As a result, all bond terms within a given $\hat{H}_c$ mutually commute. This decomposition allows the interaction part to be factorized as
\begin{equation}
e^{-\Delta\tau \hat{H}_{\mathrm{int}}} = \prod_{c=1}^{N_c} e^{-\Delta\tau \hat{H}_c} + \mathcal{O}(\Delta_\tau^2) = \prod_{c=1}^{N_c}( \prod_{b \in c} e^{-\Delta\tau \hat{H}_{c,b}}) + \mathcal{O}(\Delta_\tau^2),
\end{equation}
 \begin{figure}[t] 
    \centering
    \includegraphics[width=0.95\linewidth]{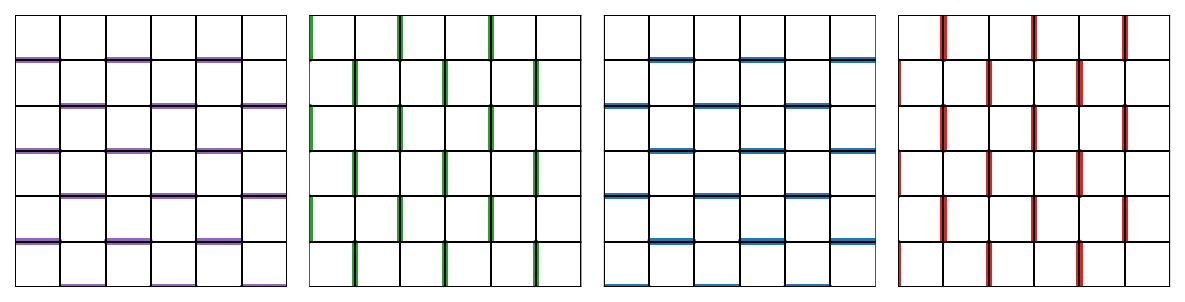}  
    \caption{ Staggered patterns used for the checkerboard decomposition.
    }
    \label{fig:S1}
\end{figure}

while preserving the Trotter error at order $\mathcal{O}(\Delta_\tau^2)$.
We then employ a general discrete Hubbard--Stratonovich (HS) transformation to decouple each bond interaction term.
\begin{equation}
e^{-\Delta_{\tau} \hat{H}_{c,b}} = \sum_{s=\pm1,\pm2}\gamma(s) e^{\alpha\eta(s) (\hat{c}^{\dagger}_{i_b,\sigma} \hat{c}_{j_b,\sigma}+\mathrm{h.c.})} + \mathcal{O}(\Delta_{\tau}^4),
\end{equation}
where the  fields $\eta(s)$ and  $\gamma(s)$ take the values
\begin{align}
\gamma(\pm 1) &= 1 + \frac{\sqrt{6}}{3}, & \gamma(\pm 2) &= 1 - \frac{\sqrt{6}}{3},\\
\eta(\pm 1) &= \pm \sqrt{2 \left(3 - \sqrt{6}\right)}, &
\eta(\pm 2) &= \pm \sqrt{2 \left(3 + \sqrt{6}\right)}.
\end{align}
As a result, the two-body term in the exponent is transformed into bilinear forms, which further decompose into independent bond-local $2\times 2$ blocks within each flavor, enabling efficient local updates.  After these steps, the fermionic degrees of freedom become quadratic and can be integrated out.
Consequently,
we can rewritten $\langle \psi_T | e^{-2\Theta \hat{H}} | \psi_T \rangle$ as a sum of configuration-dependent weights, each of which is a function of a set of auxiliary fields $\mathbf{s} = \{ s_{l,b}\}$ defined on imaginary-time slices $l$ and bonds $b$.
\begin{align}
\langle \psi_T | e^{-2\Theta \hat{H}} | \psi_T \rangle &= \sum_{\{ s_{l,b}\}}\bigg( \prod_{l,b} \gamma(s_{l,b}) \bigg) \langle \psi_T | \prod_{l=1}^{m} \Bigg[\prod_{c=1}^{Nc} \prod_{b \in c} e^{\alpha\eta(s_{l,b})( \hat{c}^{\dagger}_{i_b,\sigma} \hat{c}_{j_b,\sigma}         +\mathrm{h.c.})}         e^{-\Delta_{\tau} \hat{H_t}} \Bigg]    | \psi_T \rangle \\
& = \sum_{\{ s_{l,b}\}}\bigg( \prod_{l,b} \gamma(s_{l,b}) \bigg) \langle \psi_T |  \hat{U}_{\mathbf{s}}(2\Theta, 0) | \psi_T \rangle \\
& = \sum_{\{s_{l,b}\}}  \bigg( \prod_{l,b} \gamma(s_{l,b}) \bigg) \mathrm{det} [P^{\dagger}  B_{\mathbf{s}}(2\Theta,0)  P]\\
& = \sum_{\{s_{l,b}\}} W_{\mathbf{s}}
\end{align}
In the above expression, $|P\rangle$ denotes the Slater determinant representation of the trial wave function $|\psi_T\rangle$. $\hat{U}_{\mathbf{s}}(\tau_2,\tau_1)$ and $B_{\mathbf{s}}(\tau_2,\tau_1)$ represents the time-ordered imaginary time propagators for a given auxiliary-field configuration $\mathbf{s}$, with $\tau_1 = l_1 \Delta\tau$ and $\tau_2 = l_2 \Delta\tau$($l_1 \leq l_2$), defined as follows:
\begin{equation}
    \hat{U}_{\mathbf{s}}(\tau_2,\tau_1) = \prod_{l_1+1}^{l_2} \ \Bigg[\prod_{c=1}^{Nc} \prod_{b \in c} e^{\alpha\eta(s_{l,b})( \hat{c}^{\dagger}_{i_b,\sigma} \hat{c}_{j_b,\sigma}         +\mathrm{h.c.})}         e^{-\Delta_{\tau} \hat{H_t}} \Bigg]
\end{equation}
\begin{equation}
    B_{\mathbf{s}}(\tau_2,\tau_1) = \prod_{l_1+1}^{l_2} \ \Bigg[\prod_{c=1}^{Nc} \prod_{b \in c} e^{\alpha\eta(s_{l,b}) K_b  }       e^{-\Delta_{\tau} T} \Bigg]
\end{equation}
Here, $T$ denotes the hopping matrix corresponding to  $\hat{H}_t$, while $K_b$ is the bond operator acting on bond $b$, with nonzero elements only at $(i_b, j_b)$ and $(j_b, i_b)$, i.e., $(K_b)_{i_b,j_b} = (K_b)_{j_b,i_b} = 1$. 

Now, the expectation value of an observable $\hat{O}$ can also be rewritten as a weighted average over auxiliary-field configurations $\{s_{l,b}\}$:
\begin{equation}
    \langle \hat{O} \rangle = \frac{ \sum_{\mathbf{s}} W_{\mathbf{s} }     \langle \hat{O} \rangle_{\mathbf{s}}} { \sum_{\mathbf{s}} W_{\mathbf{s}} }.
\end{equation}
Here, $\langle \hat{O} \rangle_{\mathbf{s}}$ denotes the expectation value of $\hat{O}$ evaluated for a fixed auxiliary-field configuration $\mathbf{s}$:
\begin{equation}
    \langle \hat{O} \rangle_{\mathbf{s}}  = \frac{ \langle \psi_T|  \hat{U}(2\Theta,\Theta)\hat{O}\hat{U}(\Theta,0)|\psi_T\rangle             }{\langle \psi_T| \hat{U}(2\Theta,0)|\psi_T\rangle}  .
\end{equation}
Since each flavor sector is identical and real, the total weight $W_{\mathbf{s}}$ takes the form of an $N$-th power of a real number. Consequently, for any even $N$ and at arbitrary filling, the model is free from the sign problem.  Thus, we can construct a Markov chain that samples the auxiliary–field configurations according to their weights
$W_{\mathbf{s}}$, using the standard Metropolis update scheme.

In DQMC, the most important quantity to compute is the equal-time Green’s function $\big( G_{\mathbf{s}}(\tau,\tau) \big)_{ij} = \langle \hat{c}_i \hat{c}_j^{\dagger}  \rangle_{\mathbf{s}}$, with $\tau = \Theta$ corresponding to measurements at the middle of the projection. Its explicit expression is given by:
\begin{equation}
   G_{\mathbf{s}}(\tau,\tau) =  1 - B_{\mathbf{s}}(\tau,0)P\big( P^{\dagger} B_{\mathbf{s}}(2\Theta,0)P   \big)^{-1} P^{\dagger} B_{\mathbf{s}}(2\Theta,\tau)
\end{equation}
For a given auxiliary-field configuration, the fermionic Hamiltonian becomes quadratic and thus describes noninteracting fermions. As a result, once the equal-time Green’s function is obtained, any observable, such as equal-time correlation functions, can be evaluated using Wick’s theorem.
Green’s functions at different time slices are related through the imaginary-time propagators. Specifically, one has
\begin{subequations}
\begin{align}
    G_{\mathbf{s}}(\tau + \Delta_{\tau},\tau + \Delta_{\tau}) & = B_{\mathbf{s}}(\tau+\Delta_{\tau},\tau) G_{\mathbf{s}}(\tau,\tau) B_{\mathbf{s}}^{-1}(\tau + \Delta_{\tau},\tau) \\
    G_{\mathbf{s}}(\tau - \Delta_{\tau},\tau - \Delta_{\tau}) & = B_{\mathbf{s}}^{-1}(\tau,\tau - \Delta_{\tau}) G_{\mathbf{s}}(\tau,\tau) B_{\mathbf{s}}(\tau,\tau - \Delta_{\tau})     
\end{align}
\end{subequations}
To compute unequal-time correlation functions, we need the time-displaced Green’s function $G_{\mathbf{s}}(\tau_1,\tau_2)$, defined as follows:
\begin{equation}
    G_{\mathbf{s}}(\tau_1,\tau_2)_{ij} = \langle \hat{T} \hat{c}_i(\tau_1) \hat{c}_j^{\dagger}(\tau_2) \rangle_{\mathbf{s}}
    =\begin{cases}
\langle c_i(\tau_1) c_j^\dagger(\tau_2) \rangle_{\mathbf{s}}, & \tau_1 \ge \tau_2, \\
-\langle c_j^\dagger(\tau_2) c_i(\tau_1) \rangle_{\mathbf{s}}, & \tau_1 < \tau_2.
\end{cases}
\end{equation}
The time-displaced Green’s function matrices can be obtained by propagating the equal-time Green’s function forward and backward in imaginary time using the relations:
\begin{subequations}
\begin{align}
G_\mathbf{s}(\tau_1,\tau_2) &= B_{\mathbf{s}}(\tau_1,\tau_2)G_{\mathbf{s}}(\tau_2,\tau_2), \tau_1 > \tau_2\\
G_\mathbf{s}(\tau_1,\tau_2) &=  -(1-G_{\mathbf{s}}(\tau_1,\tau_1))B_{\mathbf{s}}^{-1}(\tau_2,\tau_1), \tau_1 < \tau_2
\end{align}
\end{subequations}

The formulation of FTQMC largely follows that of PQMC, with only a few differences. We briefly summarize the key modifications below. Firstly, in the grand-canonical ensemble, the chemical potential is taken into account by redefining the hopping Hamiltonian $\hat{H}_t$ to include the term $-\mu \hat{N}$.  Secondly, the denominator of Eq.~\eqref{eqft}, i.e., the partition function, is expressed as a sum over configuration-dependent weights:
\begin{align}
    Z &= \sum_{\{s_{l,b}\}} \bigg( \prod_{l,b} \gamma(s_{l,b}) \bigg)   \mathrm{Tr} \Bigg( \prod_{l=1}^{m} \Bigg[\prod_{c=1}^{Nc} \prod_{b \in c} e^{\alpha\eta(s_{l,b})( \hat{c}^{\dagger}_{i_b,\sigma} \hat{c}_{j_b,\sigma}         +\mathrm{h.c.})}         e^{-\Delta_{\tau} \hat{H_t}} \Bigg] \Bigg)\\
    & = \sum_{\{ s_{l,b}\}} \bigg( \prod_{l,b} \gamma(s_{l,b}) \bigg) \mathrm{Tr} \big( \hat{U}_{\mathbf{s}}(\beta,0) \big)\\
    & = \sum_{\{ s_{l,b}\}} \bigg( \prod_{l,b} \gamma(s_{l,b}) \bigg) \mathrm{det}\big(1 + B_{\mathbf{s}}(\beta,0)\big)\\
    & = \sum_{\{ s_{l,b} \}} W_{\mathbf{s}},
\end{align}
where $m \Delta_{\tau} = \beta$. We can still rewrite $\langle \hat{O} \rangle$ as $\langle \hat{O} \rangle = \frac{\sum_{\mathbf{s}} W_{\mathbf{s}} \langle \hat{O} \rangle_{\mathbf{s}} }{\sum_{\mathbf{s}} W_{\mathbf{s}}}$, but now $\langle \hat{O} \rangle_\mathbf{s}$ is:
\begin{equation}
    \langle \hat{O} \rangle_{\mathbf{s}} = \frac{\mathrm{Tr}(\hat{U}_{\mathbf{s}}(\beta,0) \hat{O} )}{\mathrm{Tr}(\hat{U}_{\mathbf{s}}(\beta,0))}.
\end{equation}
The equal-time Green’s function $G_{\mathbf{s}}(\tau,\tau)$ is now given by:
\begin{equation}
        G_{\mathbf{s}}(\tau,\tau) = (1 + B_{\mathbf{s}}(\tau,0)  B_{\mathbf{s}}(\beta,\tau)  )^{-1}.
\end{equation}
The computation of unequal-time Green’s functions proceeds in the same way as in PQMC. Observables can still be  evaluated using Wick’s theorem.

Our simulations employ both projector QMC (PQMC) for ground-state properties and finite-temperature QMC (FTQMC) for thermodynamic analysis. To ensure the accuracy of the Trotter decomposition, the imaginary-time step is set adaptively: we use $\Delta_{\tau} = 0.05$ in the weak-coupling regime and a smaller step of $\Delta_{\tau} = 0.025$ at strong coupling to accurately resolve the larger energy scales. This choice keeps Trotter errors negligible in all simulations. For the PQMC calculations, a projection time of $\Theta = 20-40$. The trial wave function, $|\psi_T\rangle$ , is chosen as the ground state of the non-interacting hopping Hamiltonian $\hat{H}_t$. In the FTQMC simulations, In FTQMC, for each set of parameters $(T, J)$, the chemical potential $\mu$ is tuned following Ref.~\cite{mu_tune} to reach the target hole doping $\delta = -0.15$.

To obtain reliable statistics, we employ 56--112 independent Markov chains in both PQMC and FTQMC simulations. For each Markov chain, we first perform a thermalization stage to reach equilibrium, followed by 250--500 measurement sweeps. In the zero-temperature strong-coupling regime, where charge-$4e$ fluctuations are significantly enhanced and lead to larger statistical uncertainties, we increase the number of independent Markov chains up to 224 and perform additional sampling to improve statistical accuracy.
In FTQMC simulations, lower temperatures and stronger coupling require longer equilibration times for the particle density to converge to the target value. 

\section{Section~II.~Finite-Size Scaling and Thermodynamic Limit Extrapolation}
In this section, we perform a finite-size scaling analysis of the charge-$2e$ and charge-$4e$ superconducting structure factors evaluated at the peak momentum $\mathbf{q}=(0,0)$, and extrapolate them to the thermodynamic limit, respectively. The structure factors are defined as:
\begin{align}
    S_{2e}(\mathbf{q}=(0,0),L) &= \frac{1}{L^4}  \sum_{ij} \langle \hat{\Delta}^{\dagger}_{2e}(i) \hat{\Delta}_{2e}(j) \rangle,\\
    S_{4e}(\mathbf{q}=(0,0),L) &= \frac{1}{L^4} \sum_{ij} \langle \hat{\Delta}_{4e}^{\dagger}(i) \hat{\Delta}_{4e}(j) \rangle.
\end{align}

Here, $\hat{\Delta}^{\dagger}_{2e}(i) = \hat{c}^{\dagger}_{i\alpha} \hat{c}^{\dagger}_{i\beta}$ denotes the local pairing operator that creates a charge-$2e$ pair of fermions with different flavors $\alpha$ and $\beta$ on site $i$. Without loss of generality, we take $\hat{\Delta}^{\dagger}_{2e}(i) = \hat{c}^{\dagger}_{i1} \hat{c}^{\dagger}_{i2}$. The charge-$4e$ pairing operator is defined as $\hat{\Delta}^{\dagger}_{4e}(i) = \hat{c}^{\dagger}_{i1} \hat{c}^{\dagger}_{i2} \hat{c}^{\dagger}_{i3} \hat{c}^{\dagger}_{i4}$, which creates a bound state carrying total charge $4e$.

\begin{figure}[htbp] 
    \centering
    \includegraphics[width=1.0\linewidth]{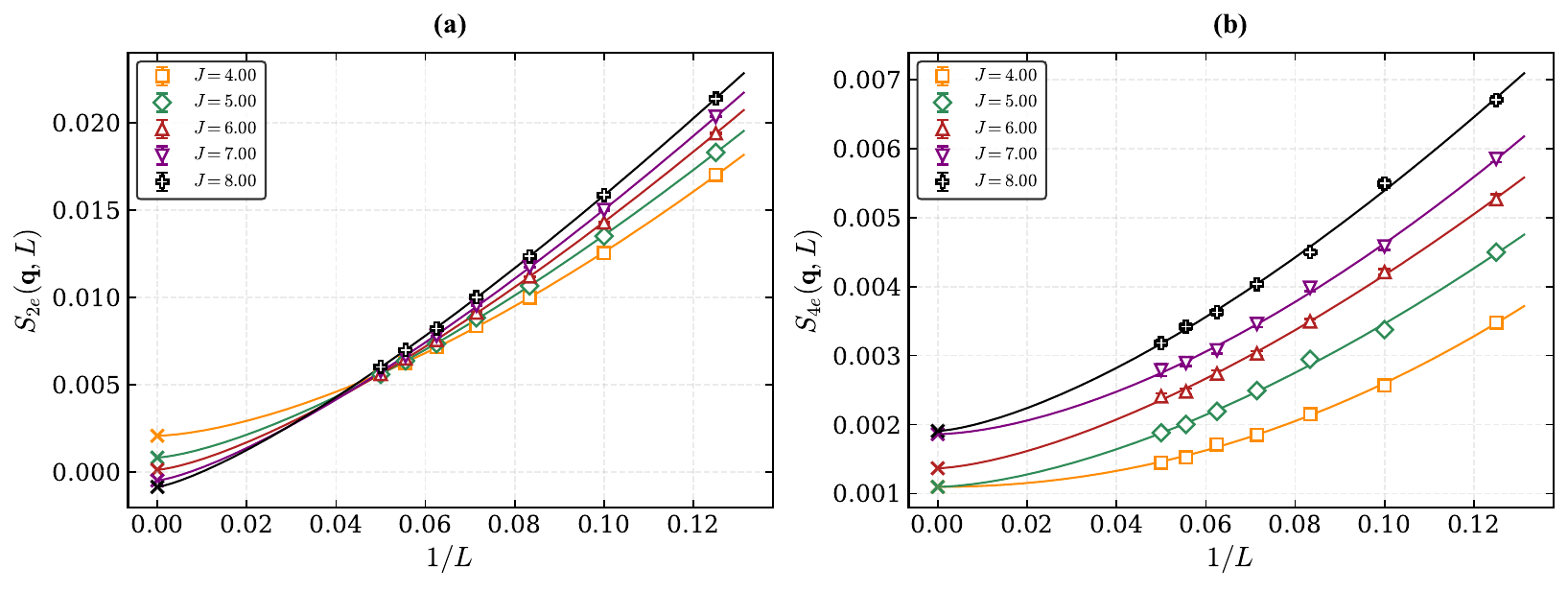}  
    \caption{
    Finite-size scaling of the charge-$2e$ and charge-$4e$ superconducting structure factors $S_{2e}(\mathbf{q}=(0,0),L)$ and $S_{4e}(\mathbf{q}=(0,0),L)$ versus $1/L$, respectively. The data are fitted to the form $a/L^b + c$, where the thermodynamic-limit value is given by the extrapolated parameter $c$.
    }
    \label{fig:S2}
\end{figure}

We then present the extrapolation results for various values of $J$. As shown in Fig.~\ref{fig:S2}(a), in the weak-coupling regime, the extrapolated value of $S_{2e}(\mathbf{q},L)$ remains finite, indicating the presence of long-range charge-$2e$ superconducting order. In contrast, in the strong-coupling regime, $S_{2e}$ extrapolates to zero, suggesting the absence of such order. Meanwhile, the extrapolated $S_{4e}(\mathbf{q},L)$ remains finite in the strong-coupling regime, indicating the emergence of a robust charge-$4e$ superconducting phase at large $J$, where charge-$4e$ correlations dominate despite the suppression of charge-$2e$ order.

\section{Section~III.~ The details for the calculation of superfluid stiffness }
We consider the general $\mathrm{SU}(N)$-symmetric SSH model defined on a square lattice:
\begin{equation}
\hat{H} = -t \sum_{\langle ij \rangle,\sigma} \left( \hat{c}^{\dagger}_{i\sigma} \hat{c}_{j\sigma} + \mathrm{h.c.} \right)
- \frac{J}{2N} \sum_{\langle ij \rangle} \left( \hat{c}^{\dagger}_{i\sigma} \hat{c}_{j\sigma} + \mathrm{h.c.} \right)^2
- \mu \sum_{i\alpha} \hat{c}^{\dagger}_{i\sigma} \hat{c}_{i\sigma},
\end{equation}
where $\mu = 0$ corresponds to half filling. Although we focus on the $\mathrm{SU}(4)$ case in the main text, the following derivation applies to general $\mathrm{SU}(N)$.

To derive the current operators, we first introduce the following auxiliary operators:
\begin{align}
   \hat{K}_{i,a} &= \,\,\,\,\,\, \hat{c}^{\dagger}_{i+a,\sigma} \hat{c}_{i,\sigma} + \hat{c}^{\dagger}_{i,\sigma} \hat{c}_{i+a,\sigma}\: \,\,\,, \\
\hat{J}_{i,a} &= i \left( \hat{c}^{\dagger}_{i+a,\sigma} \hat{c}_{i,\sigma} - \hat{c}^{\dagger}_{i,\sigma} \hat{c}_{i+a,\sigma} \right),
\end{align}

where $a = x, y$. In terms of these operators, the Hamiltonian can be rewritten as
\begin{equation}
\hat{H} = -t \sum_{i,a} \hat{K}_{i,a}
- \frac{J}{2N} \sum_{i,a} \hat{K}_{i,a}^2
- \mu \sum_i \hat{n}_i,
\end{equation}
where $\hat{n}_i = \sum_{\alpha} \hat{c}^{\dagger}_{i\sigma} \hat{c}_{i\sigma}$.

We now introduce a gauge field $A_{ix}$ along the $x$-direction via minimal coupling. The hopping term is modified by a Peierls phase:
\begin{align}
    \hat{K}_{ix}(A_{ix}) &= \;\;\;\,e^{iA_{ix}} \hat{c}^{\dagger}_{i+x,\sigma} \hat{c}_{i,\sigma}
+ e^{-iA_{ix}} \hat{c}^{\dagger}_{i,\sigma} \hat{c}_{i+x,\sigma}\;\;,\\
\hat{J}_{ix}(A_{ix}) &= i \left(
e^{iA_{ix}} \hat{c}^{\dagger}_{i+x,\sigma} \hat{c}_{i,\sigma}
- e^{-iA_{ix}} \hat{c}^{\dagger}_{i,\sigma} \hat{c}_{i+x,\sigma}
\right).
\end{align}
The current operator and diamagnetic contribution are defined as
\begin{equation}
\hat{J}^{P}_{ix} = - \left. \frac{\partial \hat{H}(A_{ix})}{\partial A_{ix}} \right|_{A_{ix}=0}, \qquad
\hat{K}_{\mathrm{dia}} = - \left. \frac{\partial^2 \hat{H}(A_{ix})}{\partial A_{ix}^2} \right|_{A_{ix}=0}.
\end{equation}

Using the relations
\begin{equation}
\frac{\partial \hat{K}_{ix}(A_{ix})}{\partial A_{ix}} = \hat{J}_{ix}(A_{ix}), \qquad
\frac{\partial \hat{J}_{ix}(A_{ix})}{\partial A_{ix}} = -\hat{K}_{ix}(A_{ix}),
\end{equation}
we obtain the current operator:
\begin{equation}
\hat{J}^{P}_{ix}
= t \hat{J}_{ix}
+ \frac{J}{2N} \left( \hat{K}_{ix} \hat{J}_{ix} + \hat{J}_{ix} \hat{K}_{ix} \right).
\end{equation}

Explicitly, it can be written as
\begin{equation}
\hat{J}^{P}_{ix}
= it \left( \hat{c}^{\dagger}_{i+x,\sigma} \hat{c}_{i,\sigma}
- \hat{c}^{\dagger}_{i,\sigma} \hat{c}_{i+x,\sigma} \right)
+ \frac{iJ}{N} \left[
\left( \hat{c}^{\dagger}_{i+x,\sigma} \hat{c}_{i,\sigma} \right)^2
- \left( \hat{c}^{\dagger}_{i,\sigma} \hat{c}_{i+x,\sigma} \right)^2
\right].
\end{equation}

This expression can be further rewritten in terms of single-particle hopping and pair-hopping processes:
\begin{equation}
\hat{J}^{P}_{ix}
= it \left( \hat{c}^{\dagger}_{i+x,\sigma} \hat{c}_{i,\sigma}
- \hat{c}^{\dagger}_{i,\sigma} \hat{c}_{i+x,\sigma} \right)
+ \frac{iJ}{N} \sum_{\alpha \beta}
\left(
\hat{\Delta}^{\dagger}_{i+x,\alpha\beta} \hat{\Delta}_{i,\alpha\beta}
- \hat{\Delta}^{\dagger}_{i,\alpha\beta} \hat{\Delta}_{i+x,\alpha\beta}
\right),
\end{equation}
where the pair creation operator is defined as
\begin{equation}
\hat{\Delta}^{\dagger}_{i,\alpha\beta}
= \hat{c}^{\dagger}_{i\alpha} \hat{c}^{\dagger}_{i\beta}.
\end{equation}

Similarly, the diamagnetic term reads
\begin{equation}
\hat{K}_{\mathrm{dia}} =
-t \hat{K}_{ix}
+ \frac{J}{N} \left( \hat{J}_{ix}^2 - \hat{K}_{ix}^2 \right)
= -t \left( \hat{c}^{\dagger}_{i+x,\sigma} \hat{c}_{i,\sigma}
+ \hat{c}^{\dagger}_{i,\sigma} \hat{c}_{i+x,\sigma} \right)
- \frac{2J}{N} \left[
\left( \hat{c}^{\dagger}_{i+x,\sigma} \hat{c}_{i,\sigma} \right)^2
+ \left( \hat{c}^{\dagger}_{i,\sigma} \hat{c}_{i+x,\sigma} \right)^2
\right].
\end{equation}
$\hat{K}_{\mathrm{dia}}$ can also be decomposed into single-particle hopping and pair-hopping contributions.
\begin{equation}
    \hat{K}_{\mathrm{dia}} = 
 -t \left( \hat{c}^{\dagger}_{i+x,\sigma} \hat{c}_{i,\sigma}
+ \hat{c}^{\dagger}_{i,\sigma} \hat{c}_{i+x,\sigma} \right)
- \frac{2J}{N} \sum_{\alpha\beta} \left(
\hat{\Delta}^{\dagger}_{i+x,\alpha\beta} \hat{\Delta}_{i,\alpha\beta}
+ \hat{\Delta}^{\dagger}_{i,\alpha\beta} \hat{\Delta}_{i+x,\alpha\beta}
\right)
\end{equation}
To evaluate the superfluid stiffness, it is crucial to compute the paramagnetic current--current correlation function, defined as
\begin{equation}
\Lambda_{xx}(\mathbf{q}, i\omega_m)
= \frac{1}{L^2} \sum_{i,j} \int_{0}^{\beta} d\tau \, e^{i\omega_m \tau} e^{-i \mathbf{q} \cdot (\mathbf{R}_i-\mathbf{R}_j)}
\left\langle \hat{J}^P_{x}(\mathbf{R}_i,\tau)\,
\hat{J}^P_{x}(\mathbf{R}_j,0) \right\rangle .
\end{equation}
The superfluid stiffness $\rho_s$ is then given by
\begin{equation}
\rho_s = \frac{1}{4} \left[ -\langle \hat{K}_{\mathrm{dia}} \rangle
- \Lambda_{xx}(q_x=0,q_y \rightarrow 0\, ,i\omega_m = 0) \right].
\end{equation}
Here we take the static, transverse, long-wavelength limit of the paramagnetic response, corresponding to the Coulomb gauge condition $\nabla \cdot \mathbf{A} = 0$, i.e., $\mathbf{q} \cdot \mathbf{A} = 0$.
The $U(1)$ gauge invariance ensures that in the static longitudinal long-wavelength limit, the paramagnetic response exactly cancels the diamagnetic contribution, i.e.,
\begin{equation}
\Lambda_{xx}(q_x \rightarrow 0,q_y=0, i\omega_m=0)
= - \langle \hat{K}_{\mathrm{dia}} \rangle .
\end{equation}
Therefore, the superfluid stiffness $\rho_s$ can also be expressed as the difference between the longitudinal and transverse responses, i.e.,
\begin{align}
\rho_s &= \frac{1}{4} \left[ \Lambda_{xx}(q_x \rightarrow0,q_y=0, i\omega_m=0)
- \Lambda_{xx}(q_x=0,q_y \rightarrow0, i\omega_m=0) \right]\\
&=\frac{1}{4} (\Lambda^{\mathrm{L}}  - \Lambda^{\mathrm{T}}).
\end{align}

\section{Section~IV.~Order-Parameter Manifold and Half-Vortex-Driven BKT Transition}

In this section, we analyze how the charge-$2e$ pairing order parameter transforms under the symmetry group of the model introduced in the main text. We then determine the corresponding order-parameter manifold associated with spontaneous symmetry breaking. We show that this manifold contains the identification $\theta \sim \theta + \pi$, which allows for thermally excited half-vortex defects. We interpret this as a finite-temperature charge-$4e$ phase that supports half-vortices, and the associated BKT transition is characterized by the universal superfluid-stiffness jump
\begin{equation}
\rho_s(T_c^-)=\frac{8T_c}{\pi}.
\end{equation}

Away from half filling, the microscopic Hamiltonian possesses a global $U(1)$ symmetry associated with particle-number conservation, together with an $SU(4)$ flavor symmetry under which the four fermionic flavors are equivalent. The full symmetry group is therefore
\begin{equation}
G=\frac{U(1)\times SU(4)}{\mathbb{Z}_4}\simeq U(4),
\end{equation}
where the quotient accounts for the common $\mathbb{Z}_4$ center.

The local pairing operator $\hat c_{ia}\hat c_{ib}$ is antisymmetric in the flavor indices because of the fermionic anticommutation relations. It therefore transforms in the rank-2 antisymmetric representation of $SU(4)$,
\begin{equation}
4\otimes 4 = 6 \oplus 10,
\end{equation}
with the pairing channel corresponding to the six-dimensional antisymmetric sector. This representation is equivalent to the vector representation of $SO(6)$, consistent with the isomorphism $SU(4)/\mathbb{Z}_2 \simeq SO(6)$.

It is convenient to organize the order parameter into an antisymmetric matrix $\Delta_{ab}=-\Delta_{ba}$, such that the pairing field is written as $\hat\Delta_i=\hat c_{ia}\Delta_{ab}\hat c_{ib}$. Under an $SU(4)$ transformation $\hat c \to U\hat c$, the order parameter transforms as
\begin{equation}
\Delta \rightarrow U^{T}\Delta U.
\end{equation}
Because the center elements $\pm \mathbb{I}\in SU(4)$ act identically on $\Delta$, the induced action on the order parameter is two-to-one, again reflecting the identification $SU(4)/\mathbb{Z}_2 \simeq SO(6)$.

In the symmetry-broken ground state, one may choose a reference configuration
\begin{equation}
\Delta_0=\sigma_0\otimes i\sigma_y=
\begin{pmatrix}
0&1&0&0\\
-1&0&0&0\\
0&0&0&1\\
0&0&-1&0
\end{pmatrix},
\end{equation}
which pairs all four fermion flavors and opens a spectral gap. The subgroup leaving $\Delta_0$ invariant is $Sp(4)$, so the degeneracy manifold is
\begin{equation}
\mathcal{M}_0=\frac{SU(4)}{Sp(4)}=\frac{SO(6)}{SO(5)}\simeq S^5.
\end{equation}

After including the global $U(1)$ phase, one might naively expect the order-parameter manifold to be $S^5 \times U(1)$. However, there exist combined $SU(4)$ and $U(1)$ transformations for which
\begin{equation}
A \in SU(4), \, \, \,A^T \Delta_0 A = -\Delta_0,
\label{eq:Aminus}
\end{equation}
so that the minus sign can be absorbed by a $U(1)$ phase shift $e^{i\pi}$. One representative example is
\begin{equation}
A_0 = \sigma_0 \otimes \sigma_x,
\end{equation}
and the complete set of such transformations is $A = Sp(4)\,A_0$. Consequently, the true order-parameter manifold is
\begin{equation}
\mathcal{M}=\frac{S^5 \times U(1)}{\mathbb{Z}_2},
\end{equation}
where the $\mathbb{Z}_2$ quotient identifies $\theta$ and $\theta+\pi$ in the $U(1)$ sector.

This $\mathbb{Z}_2$ identification allows for half-vortex topological defects. At finite temperature, the Mermin--Wagner theorem forbids true long-range order of the charge-$2e$ order parameter, since it transforms nontrivially under the continuous $SO(6)$ symmetry. By contrast, the on-site charge-$4e$ order parameter is an $SU(4)$ singlet and therefore does not break the flavor symmetry, so it may still exhibit quasi-long-range order at finite temperature. We thus interpret the finite-temperature phase as a charge-$4e$ phase supporting half-vortices. Correspondingly, the BKT transition is determined by the intersection of the superfluid stiffness $\rho_s(T)$ with the line $8T/\pi$.

More generally, one may ask whether there exist transformations satisfying
\begin{equation}
A^{T}\Delta_0 A = e^{i\alpha}\Delta_0
\end{equation}
for phases other than $\alpha=0$ and $\pi$. Using the Pfaffian identities for antisymmetric matrices,
\begin{equation}
\mathrm{pf}(BAB^T)=\det(B)\,\mathrm{pf}(A), 
\qquad
\mathrm{pf}(\lambda A)=\lambda^2 \mathrm{pf}(A),
\end{equation}
and applying $\mathrm{pf}$ to both sides of the above equation, one obtains
\begin{equation}
e^{i2\alpha}=1,
\end{equation}
which restricts $\alpha$ to $0$ and $\pi$.

For an $SU(2N)$ generalization, the corresponding condition becomes
\begin{equation}
e^{iN\alpha}=1,
\end{equation}
so that $\alpha=2\pi k/N$ with $k=0,1,\dots,N-1$. In that case, fractional vortices with vorticity $1/N$ may emerge, and the KT transition is expected to occur at the intersection of the superfluid stiffness with the line $2N^2T/\pi$.

\section{Section~V.~ BKT Transition and spectral properties in the Charge-$\scalebox{1.2}{$4e$}$ Phase}

\begin{figure}[htbp] 
    \centering
    \includegraphics[width=0.95\linewidth]{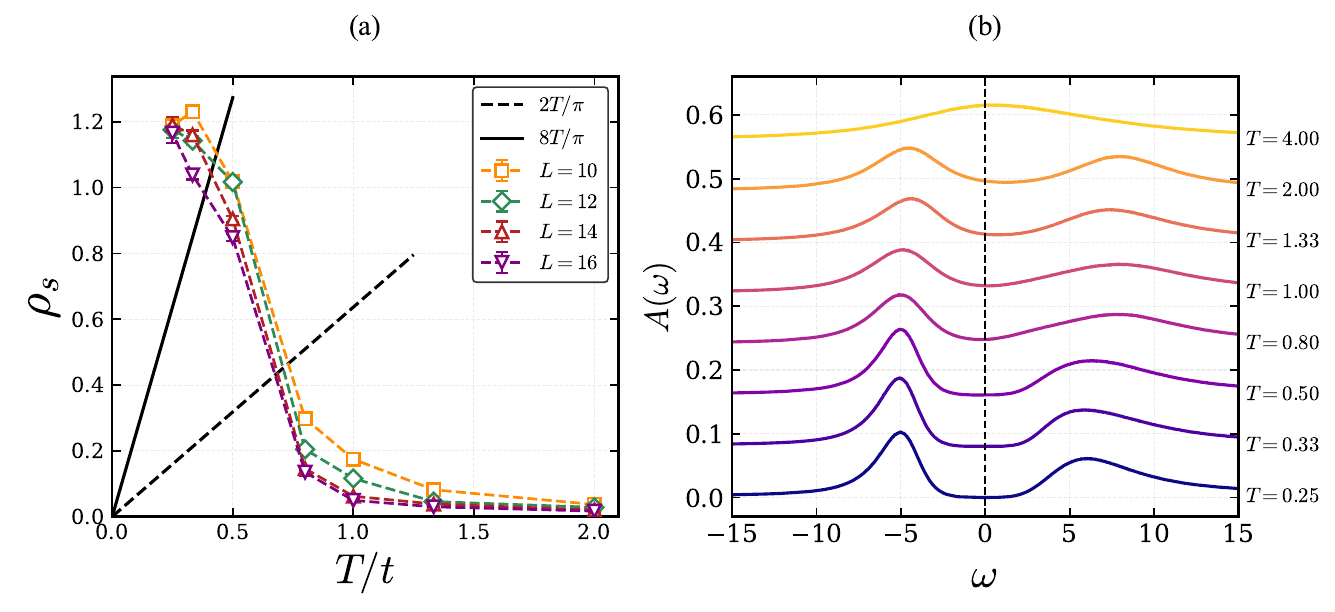}  
    \caption{  Results for the charge-$4e$ superconducting phase at $J=8.0$.
(a) Superfluid stiffness $\rho_s$ as a function of temperature $T$. The BKT transition temperature is identified from the intersection of $\rho_s(T)$ with the universal line $8T/\pi$.
(b) Finite-temperature single-particle spectral function, where the persistence of the pseudo-gap feature is clearly observed.
    }
    \label{fig:S3}
\end{figure}

In this section, we provide further evidence for the BKT transition in the charge-$4e$ superconducting phase by analyzing the superfluid stiffness and the single-particle spectral function. Complementing the $J=4$ and $J=6$ data presented in the main text, here we focus on $J=8$ in the strong-coupling regime, where the ground state is a charge-$4e$ superconductor. Figure~\ref{fig:S3}(a) displays the superfluid stiffness $\rho_s$ as a function of temperature $T$, confirming that the transition point is indeed determined by the intersection of $\rho_s(T)$ with the $8T/\pi$ line. Furthermore, Fig.~\ref{fig:S3}(b) shows the finite-temperature single-particle spectral function $A(\omega)$. Below the transition temperature $T_c \approx 0.4$, $A(\omega)$ exhibits a full gap accompanied by coherence peaks. Notably, this gap feature persists over a broad temperature range above $T_c$. This pseudogap temperature window in the strong-coupling charge-$4e$ phase is significantly wider than in the weak-coupling regime, where the ground state is a charge-$2e$ superconductor.

\section{Section~VI.~ Stochastic Analytic Continuation}

In the main text, we have demonstrated the presence of a pseudogap in the model, where the single-particle spectral function $
A(\omega) = \frac{1}{L^2} \sum_{\mathbf{k}} A(\mathbf{k}, \omega)
$
 remains gapped above the critical temperature over a finite temperature range. 

In quantum Monte Carlo simulations, the spectral function is not directly accessible in real frequency. Instead, calculations are performed in imaginary time, yielding the imaginary-time Green's function
\begin{equation}
    G(\mathbf{k}, \tau) = \langle c_{\mathbf{k}}(\tau) c^{\dagger}_{\mathbf{k}}(0) \rangle.
\end{equation}
The connection between the imaginary-time Green's function and the real-frequency spectral function is given by the spectral representation
\begin{equation}
G(\mathbf{k}, \tau) = \int_{-\infty}^{\infty} d\omega \, K(\tau, \omega)\, A(\mathbf{k}, \omega),
\end{equation}
where the kernel is
\begin{equation}
    K(\tau, \omega) = \frac{e^{-\tau \omega}}{1 + e^{-\beta \omega}},
\end{equation}
and $A(\mathbf{k},\omega)$ satisfies the normalization condition $\int_{-\infty}^{\infty} d\omega \, A(\mathbf{k}, \omega) = 1$.
The extraction of $A(\mathbf{k}, \omega)$ (and consequently $A(\omega)$) requires inverting this integral transform, which is an ill-posed problem. To overcome this difficulty, we employ the well-established stochastic analytical continuation method~\cite{shao2023SAC}.

\end{document}